%% file: preprint.tex
\documentstyle[epsfig,12pt]{article}
\begin{document}
\thispagestyle{empty}
\centerline{ \large EUROPEAN ORGANIZATION FOR NUCLEAR RESEARCH (CERN)} 
\vspace{0.5cm}
\begin{flushright}
CERN-EP/2001-047 \\
29 June 2001 \\
\end{flushright}
\vspace{3.5cm}
\boldmath
\vspace{1cm}
\begin{center}
{\huge \bf Measurement of $A_{\rm FB}^{b}$ using \\ 
  \vspace{1mm} Inclusive $b$-hadron Decays  }
\end{center}
\unboldmath
\vspace{1 cm}
\begin{center}
{\Large The ALEPH collaboration }
\end{center}
\vspace{1 cm}
%
\begin{abstract}
Based on a sample of four million events collected by ALEPH from
1991 to 1995, a
measurement of the forward-backward asymmetry in $Z
\rightarrow b\bar{b}$ decays using inclusive final states
is presented.
High-performance tagging of $b\bar{b}$ events
in a wide angular range is achieved using neural network techniques.
An optimal hemisphere charge estimator is built by merging
primary and secondary vertex information, leading kaon
identification and jet charge in a neural network.  The
average charge asymmetry, the flavour
tagging efficiencies and mean $b$-hemisphere charges are measured
from data and used to extract the pole $b$ asymmetry
in the Standard Model 
\begin{eqnarray*}
\nonumber                      
 A^{0,b}_{\rm FB} & = & 0.1009 \,
 \pm \, 0.0027\, (stat) \, \pm \, 0.0012\, (syst)\, ,
\end{eqnarray*}
corresponding to a value of the effective weak
mixing angle of
\begin{center}
 ${\rm sin^2} \theta_{\rm W}^{\rm eff} \: = \: 0.23193 \pm 0.00056$\, .
\end{center}
\end{abstract}
\vspace{0.5 cm}
\begin{center}
{\large \em submitted to Eur.\ Phys.\ J.\ C  }
\end{center}

\include{authors}

\section{Introduction}
The measurements of the
\mbox{forward-backward} asymmetry of $b$-quarks from $e^+ e^- \rightarrow Z
\rightarrow b\bar{b}$ production offer the most precise
determination of the weak mixing angle at LEP. This gives a sensitive
test of Standard Model~\cite{SM} predictions of electroweak radiative
corrections and hence constrain the
allowed range of the Higgs boson mass.

The polar angle distribution of the $b$-quark in the $e^+ e^- \rightarrow
Z \rightarrow b \bar{b}$ process is forward-backward asymmetric:

\begin{eqnarray}
 \frac{d \sigma_{b\bar{b}} }{d \cos \theta_b } \: = \: \sigma_{b \bar{b}}
 \, \frac{3}{8} \,
\left( 1+\cos^2 \theta_b +
 \frac{8}{3} A^b_{\rm FB} \cos \theta_b \right) \nonumber \, ,
\end{eqnarray}
where $\theta_b$ is the angle between the incoming electron and the
outgoing $b$-quark.
In this paper an improved measurement of  $A_{\rm FB}^b$ using inclusive
$b$-hadron decays is
presented, based on the data sample recorded
by ALEPH on the $Z$ from 1991 to 1995. 
The analysis takes advantage of a reprocessing of
LEP I data with
improved charged particle tracking and of neural network techniques
making
maximum use of the available event information.

The method of selecting events containing $b\bar{b}$
quark pairs is upgraded with respect to the previous analysis \cite{lastafb}.
The $b$-tagging algorithm based on the impact parameters of charged
particle tracks is complemented with
information from displaced
secondary vertices, event shape variables and lepton
identification.  This leads to a 15\% reduction of the statistical
uncertainty.
The estimate of the $b$-quark direction
is based upon the hemisphere charge method~\cite{original},
but also incorporates information
from fast kaon tagging and separated primary and secondary vertex
charge estimators. The fraction of incorrect charge tags is
reduced by 10\% with respect to the previous measurement, implying
a 20\% further reduction of the statistical uncertainty.
Another new feature is the control of
systematic uncertainties
by use of double tag methods for both flavour and charge
tags, which yields a reduction of the systematic uncertainty
by about a factor of two.

Although lepton
identification is used to select events in the \mbox{$b$-tag}
algorithm, no use is made in the hemisphere charge method of the
lepton beyond that of an ordinary charged track in the detector.  This
reduces the correlation with the forthcoming
ALEPH $A_{\rm FB}^b$ analysis based on semileptonic final states.

\section{The method}
\label{principles}
 The measurement of $A^b_{\rm FB}$ requires knowledge of the direction of the
$b$-quark from $Z \rightarrow b\bar{b}$ decay.
The quark-antiquark
axis is estimated using the reconstructed thrust axis, the direction
of which is given throughout this paper by
a positive value for the cosine of the thrust
polar angle, $\cos \theta $.  Each
event is then divided into two
hemispheres, F and B, by a plane perpendicular to the thrust axis.
The forward hemisphere, F, is defined as the one into which the
incoming electron points. 
The \mbox{F-B} orientation of the $b$-quark is determined
on a statistical basis by estimating the hemisphere charges.

Hemisphere charges, $Q_{\mathrm{F}}$ and $Q_{\mathrm{B}}$,
are formed using a neural network
designed to optimise the separation power between $b$- and 
$\bar{b}$-quarks.  The neural network also performs well for lighter
quark flavours, albeit not in an optimal way.
A forward-backward asymmetry for flavour 
$f$ at a given value of $\cos \theta $
is then proportional to the mean charge flow,  $\langle Q_{\rm
  FB}^f \rangle$,  between forward and backward hemispheres in pure
$f\bar{f}$ events~:
\begin{eqnarray}
  \langle Q_{\mathrm FB}^f \rangle  & = & 
  \langle Q_{\mathrm F}^f  -  Q_{\mathrm B}^f \rangle \nonumber \\
  & = & \frac{1}{n^f_{\rm tot}} \,
  \left( n^f_{\rm F} \langle Q_f  -  Q_{\bar{f}} \rangle
  + n^f_{\rm B} \langle Q_{\bar{f}} - Q_f \rangle \right)
     \nonumber \\
   & = & \delta_f \, A_{\mathrm FB}^f \, \frac{8}{3} \,
  \frac{\cos \theta} {1+\cos^2 \theta} \, ,
  \label{equation-qfb diff}
\end{eqnarray}
where $n^f_{\rm F}$ ($n^f_{\rm B}$) is the number of events with the
primary quark from $Z$ decay emitted into the forward (backward) hemisphere,
and $\delta_f = \langle Q_f  -  Q_{\bar{f}} \rangle $ is
the average difference between the charges measured in the hemispheres
of the quark and anti-quark, called
the charge separation for flavour $f$.  As shown in~\cite{aleph afb},
the same sample of events used to measure $\langle Q_{\rm FB}^f \rangle$
can also be used to extract $\delta_f$.  This can be understood by
considering a single hemisphere charge measurement, $Q_f$, which can
be written as~: 
\begin{eqnarray}
  Q_f \: = \: \frac{ \delta_f }{2} \, + \, {{\cal R}}_f
  \: \: \: {\mathrm and} \: \: \:
  Q_{\bar{f}} \: = \: \frac{ \delta_{\bar{f}} }{2} \, + \,
                      {{\cal R}}_{\bar{f}}  \nonumber \, ,
\end{eqnarray}
where ${\cal R}$ represents the measurement fluctuation
due to fragmentation and
detector effects.  The product of the two hemisphere charges then
averages to~:
\begin{eqnarray}
  \langle Q_f Q_{\bar{f}} \rangle \: = \:
  \langle Q_{\mathrm F} Q_{\mathrm B} \rangle \: = \:
  \frac{ - \delta_f^2 }{4} \: + \:
  \langle {{\cal R}}_f {{\cal R}}_{\bar{f}} \rangle  \nonumber \, ,
\end{eqnarray}
defining $\delta_f = - \delta_{\bar{f}}$.
The measurement fluctuation
correlation, $\langle {{\cal R}}_f {{\cal R}}_{\bar{f}} \rangle$,
arises from effects of charge conservation, sharing a common event axis and
crossover of particles close to the hemisphere boundary.
It is then useful to
define~: 
\begin{eqnarray}
  \bar{ \delta }_f^2 & = & \sigma^2 ( Q_{\mathrm FB}^f )
                           \; - \; \sigma^2 ( Q_{\rm tot}^f ) \nonumber \\
                     & = &   \delta_f^2  \; - \;
                    4 \, \langle {{\cal R}}_f {{\cal R}}_{\bar{f}} \rangle
                       \; - \;  \langle Q_{\mathrm FB}^f \rangle^2
                       \; + \;  \langle Q_{\rm tot}^f          \rangle^2 \nonumber \\
                     & =  & \left[ \delta_f \left(1 + k_f \right) \right]^2
\, ,
  \label{equation-db from dbarb}
\end{eqnarray}
where $Q_{\rm tot}^f$ is the total charge measured in an $f\bar{f}$ event.
Thus, the observable $\bar{ \delta }_f$ is equal to $\delta_f$ to
within a correction term, $k_f$, depending on the polar angle
and taking an average
value of 9\% for heavy flavours in this analysis. Compared with the
charge separation itself, the $k_f$ correction term is less sensitive
to the details of quark fragmentation and detector resolution
\cite{aleph qfb}.
Relying on the Monte Carlo prediction of $k_f$, it is possible
to extract $\delta_f$ by fitting the quantities  $\bar{ \delta }_f$
to a range of measurements of the flavour combined $\bar \delta$
using the relation:
\begin{equation}
  \bar{\delta}^2 \: =
  \:   \sum_{f=u,d,s,c,b} \: {\cal P}_f \: \left( \bar{\delta}_f \right)^2
                \:  = 
             \: -4 \, \langle Q_{\mathrm F} Q_{\mathrm B} \rangle
                       \; - \;  \langle Q_{\mathrm FB} \rangle^2
                       \; + \;  \langle Q_{\mathrm tot} \rangle^2 
  \, , 
    \label{equation-dbar}
\end{equation}
where ${\cal P}_f$ are flavour purities of the event sample under study.
As described in Section \ref{purfit}, these purities are also measured in
the data.

For each value of $\cos \theta$ the mean charge flow, the purities
and the charge separations are measured.
The $b$-quark asymmetry is then determined according to
Equation (\ref{equation-qfb diff}), averaged over quark flavours:
\begin{equation}
   \langle Q_{\mathrm FB} (\theta ) \rangle \: = \:
 \sum_{f=u,d,s,c,b} \; {\cal P}_f (\theta ) \,
   \delta_f (\theta ) \, A^f_{\mathrm FB}( \theta ) 
  \label{equation-qfb} \, ,
\end{equation}
where
\begin{equation}
A^f_{\mathrm FB}( \theta ) \: = \: A^f_{\mathrm FB} \, \frac{8}{3}
 \, \frac{\cos \theta }{ 1 + \cos \theta^2 } \, .  
\end{equation}
Rearranging Equation~(\ref{equation-qfb}), 
\begin{equation}
  A_{\mathrm FB}^b ( \theta ) \: = \: \frac{1}{
  {\cal P}_b (\theta )  \delta_b (\theta) } \ \left( \
  \langle Q_{\mathrm FB} (\theta ) \rangle  \: - \:
          \, \sum_{f=u,d,s,c} \, {{\cal P}_f ( \theta ) }
           \delta_f (\theta ) A_{\mathrm FB}^f ( \theta )\ \right)
 \, .
  \label{eq_afb}
\end{equation}
The $b$-quark \mbox{forward-backward} asymmetry, $A_{\mathrm FB}^b$,
is extracted from a simultaneous fit to the measurements
in nine angular bins
in the range $0 < \cos \theta < 0.95$. 
A tiny, but non-zero, correction is applied
to take into account the flavour dependence of the acceptance
in each angular bin.
%
\section{The ALEPH detector}
\label{detector}

The ALEPH detector is described in detail in
\cite{ALEPH det}, and its performance
in~\cite{ALEPH perf}.
The tracking system consists of two
layers of double-sided silicon vertex-detector (VDET), an
inner tracking chamber (ITC) and a time projection chamber
(TPC). The VDET
single hit resolution is 12$\, \mu$m at normal incidence for both the
$r\phi$ and $rz$ projections and 22$\, \mu$m at maximum polar angle.
The polar angle coverage of the inner
and outer layers are $|\cos\,\theta|<0.84$ and $|\cos\,\theta|<0.69$
respectively.  The ITC provides up to 8 $r \phi$ hits at radii 16 to
26$\,$cm relative to the beam with an average resolution of 150$\,
\mu$m and has an angular coverage of $|\cos\,\theta|<0.97$.  The
TPC measures up to 21 space points per track at radii
between 38 and 171$\,$cm, with an $r\phi$ resolution of 170$\, \mu$m
and a $z$ resolution of 740$\, \mu$m and with an angular coverage of
$|\cos\,\theta|<0.96$. In addition, the TPC wire planes provide up to
338 samples of ionisation energy loss ($dE/dx$).

 Tracks are reconstructed using the  TPC,
 ITC and VDET which are immersed in a 1.5T axial
magnetic field. This provides
a transverse momentum resolution of
$\sigma(1/p_T)$ = 0.0006 (GeV/$c$)$^{-1}$ for 45 GeV muons.
Multiple scattering dominates at low momentum and adds a constant term
of 0.005 to $\sigma(p_T)/p_T$.

Outside of the  TPC, the electromagnetic
calorimeter (ECAL) consists of 45 layers of lead
interleaved with proportional wire chambers. The ECAL is used to
identify photons and electrons and gives an energy resolution
$\sigma(E)/E$ = 0.18/$\sqrt{E/\rm{GeV}} + 0.009$ for isolated particles.
The hadron
calorimeter (HCAL) is formed by the iron of the magnet return
yoke interleaved with 23 layers of streamer tubes. It is used to
measure hadronic energy and, together with two
surrounding layers of muon chambers, to identify muons.

The information from the subdetectors is combined in
an energy flow algorithm ~\cite{ALEPH perf} which gives a list
of charged and neutral track momenta.

Recently the LEP I data have been reprocessed using improved
reconstruction algorithms. In particular, the VDET hits are
distributed among tracks according to a global $\chi^2$
minimisation procedure which improves the hit association efficiency
by more than 2\%. Information from TPC wires is now used in addition
to the pad information to improve the coordinate resolution by a factor
of two in $z$, and by 30\% in $r\phi$ for low momentum tracks. Similarly
the pad information is added to the $dE/dx$ information from TPC wires,
increasing the fraction of tracks with a useful $dE/dx$ measurement
to almost 100\% and providing on average a two sigma separation
between pions and kaons with momenta above 2.5 GeV/$c$.
\section{Monte Carlo simulation}
\label{mc}

The analysis makes use of a Monte Carlo sample (MC) of 8.1 million
simulated
hadronic $Z$ decays as well as two dedicated heavy flavour
samples of 4.9 million $Z \rightarrow b \bar{b}$ decays
and 2.4 million $Z \rightarrow c\bar{c}$ decays. The simulation
is based on JETSET \cite{jetset} with parameters tuned
to reproduce inclusive
particle spectra and event shape distributions measured in hadronic
$Z$ decays \cite{jetset_tuning}.

In the MC simulation, the most relevant physics input parameters
have been adjusted, using the re-weighting technique, to the recently
measured values \cite{physpar,physin}
shown in
Table~\ref{table-physpar}. The fragmentation of
heavy quarks into hadrons is assumed to follow
the model of Peterson et al.\
\cite{peterson} with parameters tuned to match the values of the mean
heavy hadron fractions of the beam energy
reported in the table.
  \begin{table}[htbp]
    \begin{center}
      \begin{tabular}{|l|c|} \hline
Physics parameter & World average value  \\
\hline
$<x_b>$ beam energy fraction & $0.702 \pm 0.008$ \\
$<x_{c}>$ beam energy fraction & $0.484 \pm 0.008$ \\
$n_{ch}$ in $b$-hadron decay ($K^0$ and $\Lambda$ incl.)    
             &  $5.44 \pm 0.09$  \\
$B_{s}$ fraction                       & $0.100 \pm 0.012$ \\
$\Lambda_{b}$ fraction                 & $0.099 \pm 0.017$ \\
$B^+$ lifetime                       & $1.656 \pm 0.025$ ps \\
$B^0_{d}$ lifetime                     & $1.562 \pm 0.029$ ps \\
$B_{s}$ lifetime                       & $1.464 \pm 0.057$  ps \\
$b$-baryon lifetime                 & $1.208 \pm 0.051$ ps \\
$g \rightarrow {b}\bar{b}$ rate
                                          & $0.00251  \pm 0.00063$  \\
$g \rightarrow {c}\bar{c}$ rate
                                          & $0.0319  \pm 0.0046$ \\
\hline
      \end{tabular}
      \caption{List of physics input parameters \cite{physpar,physin}
        to the Monte Carlo
        simulation which are used for re-weighting.} 
      \label{table-physpar}
    \end{center}
  \end{table}


\boldmath
\section{The neural net $b$-tag}
\unboldmath
\label{nnbtag}
%
%
Due to the long lifetime and high mass of $b$-hadrons, $b$-jets have
several characteristic properties.  Six discriminating variables
are combined
using a neural network to tag $b$-quark jets.  A similar scheme is used
by ALEPH to identify $b$-quark jets in searches for neutral Higgs bosons
conducted at LEP II~\cite{Higgspapers}.

\begin{figure}[htbp]
\begin{center}\mbox{\epsfig{file=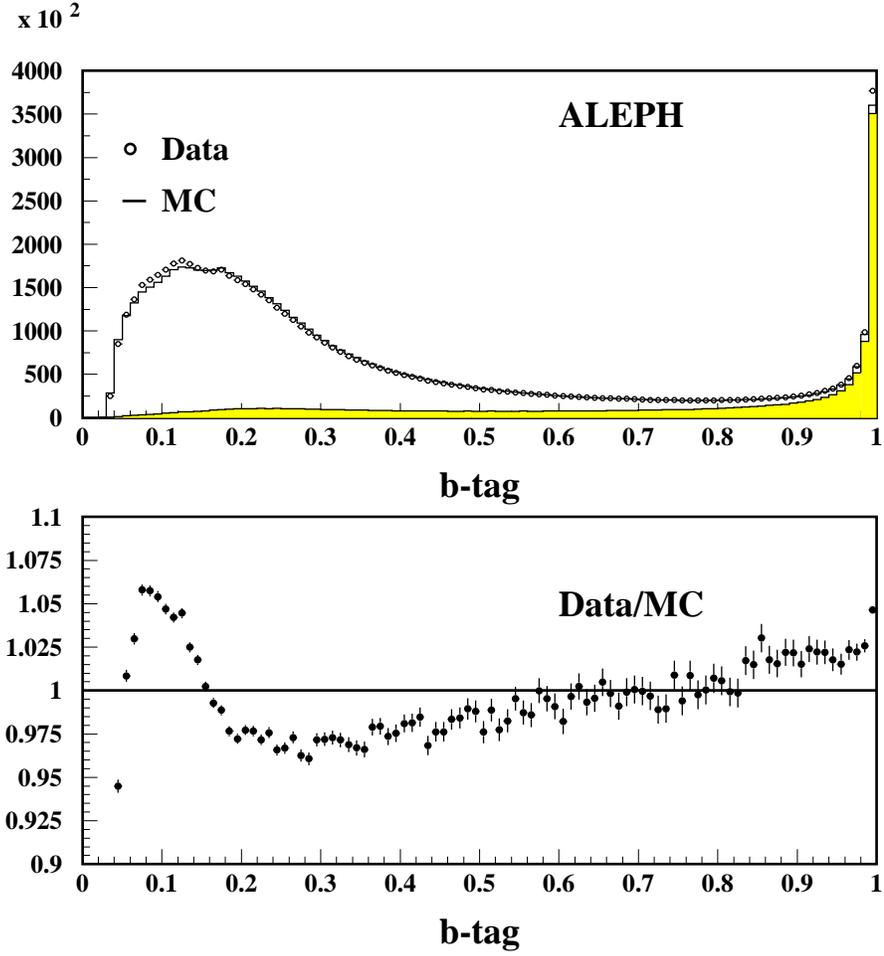,width=0.8 \textwidth}}\end{center}
\vspace{0.2 cm}
\caption{Distribution of the neural net hemisphere $b$-tags and the ratio between
data and MC. The shaded region is the distribution for $b$-hemispheres in the
MC.
\label{fig_btag}}
\end{figure}

The jets are clustered with the JADE algorithm using a $y_{cut}$
value of 0.02, and for
each jet six variables are defined:
two of them are lifetime-based; a third one is based on the
transverse momentum of identified leptons; the last three are based on
jet-shape properties.  These quantities are
\begin{enumerate}
\item $\mathrm{P}_{\mathrm{jet}}$: probability of the jet being a light
      quark ($uds$) jet based upon impact parameters of tracks in
      the jet~\cite{QIPBTAGref}.
\item $\Delta\chi^{2}_{\mathrm{svx}}$: the vertex $\chi^2$ difference 
      between assigning tracks in the jet both to the secondary and
      primary vertices compared to assigning all tracks to the
      primary vertex.  This is based upon a secondary vertex pattern
      recognition algorithm which searches for displaced vertices via
      a three-dimensional grid point search~\cite{QVSRCHref};
\item $p_{\mathrm{T}}$ of identified leptons~\cite{leptonref}
      with respect to the jet axis;
\item $\mathcal{S}^{\mathrm{b}}$: the boosted sphericity of the jet defined
      to be the sphericity of final state charged and neutral particles in
      the rest frame of the jet;
\item Multiplicity/ln(E$_{\mathrm{jet}}$): the charged and neutral particle
      multiplicity of the jet divided by the natural logarithm of the
      jet energy in GeV;
\item $\Sigma p^{2}_{\mathrm{T}}$: the sum of
      squared transverse momenta, 
      $p^{2}_{\mathrm{T}}$, of each charged or neutral particle with 
      respect to the jet axis.
\end{enumerate}

Among the six input variables, the lifetime tags have the largest weight.
However, the inclusion of the other variables increases the
$b$-quark discriminating power, especially at the most forward angles that
are not covered by the vertex detector. 
A hemisphere $b$-tag is defined as the
maximum neural net output value
among the jets in the hemisphere with energies exceeding 10 GeV.

Prior to the application of the lifetime reconstruction algorithms
in the MC,  the track parameters have been given additional smearing
and the VDET additional hit inefficiency according to a procedure
described in \cite{rblife}. This is necessary in order
to render the MC distributions of the lifetime tags
in good agreement with data. Small discrepancies still
remain and combine in the
neural network to give the differences between the distributions
of the $b$-tag
shown in Figure \ref{fig_btag}.
Deviations up to about 5\% between
data and simulation are seen in some $b$-tag
bins and for this reason the $b$-quark purity as a function of $b$-tag
is determined from the data themselves.

\section{The neural net charge tag}
A neural network is also used to determine hemisphere
charges. This technique
has been employed earlier for
CP violation studies in $B^0 / \bar{B}^0$ decays as
described
in~\cite{jpsik0s}.  In the present analysis information regarding
lepton identification
is left out from the tag.

In order to achieve high tagging performance independently
of the $b$-hadron species, momentum and polar angle, several charge estimators
are combined. These charge tags are complemented with other variables, which
do not carry information about the $b$-hadron charge in themselves,
but are correlated
with the relative tagging power of the charge estimators, thereby
allowing an optimal
combination to be achieved in the whole phase space.
The
two types of input variables are described separately in
Sections~\ref{estimators} and~\ref{deciders}. 
\subsection{Charge estimators}
\label{estimators}
Eight charge estimators are used to provide information on the
initial charge state of $b$-quark hemispheres as inclusively as
possible.  These are:
\begin{enumerate}
\item {\em The jet charge}.
This is the charge estimator used in the previous analysis \cite{lastafb}:
the weighted sum of particle charges in a hemisphere, where the weights
are the particle momenta along the thrust axis, $p_{\rm L}$,
raised to the power $\kappa$. 
Four different values of $\kappa$ are used for inputs
to the neural net: 0.0, 0.5, 1.1 and 2.2.
These focus on correspondingly
higher values of track momenta but are of course highly correlated.
Such charge
estimators work well for all species of $b$-hadrons.
\item {\em The secondary vertex charge}.
Using a topological vertexing algorithm combining information from
all charged tracks in the hemisphere \cite{jpsik0s},
a best estimate for a secondary vertex position
is obtained in each hemisphere. An estimate of the charge of this vertex
is calculated as:
\begin{equation}
Q_{\rm vtx} \: = \: \sum_{i={\rm tracks}} w_i \, q_i \, ,
\end{equation}
where $w_i$, calculated by a dedicated neural net \cite{jpsik0s},
is the probability that track $i$ comes from the
secondary vertex and $q_i$ is the track charge.
This estimator provides a high quality charge tag for $B^{\pm}$
hemispheres, and also helps to
indicate which hemispheres are more likely to contain a neutral
$B$ meson.

\item {\em The weighted primary vertex charge}. 
For  $B^0$ hemispheres the secondary vertex charge carries little
information, but in this case the fragmentation tracks close in
phase space to the $B^0$ have some correlation with the primary
$b$-quark charge.
Therefore a charge estimator,
$Q_{\rm Pvtx}$, is
calculated
according to~:
\begin{equation}
Q_{\rm Pvtx} \: = \: \sum_{i={\rm tracks}} ( 1 - w_i )
                        \, p_{\rm L}^i \, q_i \, 
                       /  \, \sum_{i={\rm tracks}} (1 - w_i) \,
                       p_{\rm L}^i \, .
\end{equation}
\item  {\em The weighted secondary vertex charge}, $Q_{\rm Svtx}$,
is calculated in a similar manner, replacing the weight
$ ( 1 - w_i )  \, p_{\rm L}^i$
by $w_i \, \left( p_{\rm L}^i \right) ^{0.3}$
with the aim of improving the tagging for both charged and neutral
$B$-decays via leading particle effects from the secondary decay.
\item {\em Fast kaon identification} is formed by another
dedicated
\mbox{sub-net}~\cite{jpsik0s} trained to identify charged kaons from
\mbox{$b$-hadron} decays. The output for the
most kaon-like particle in the hemisphere is signed by the charge
of the particle and used as a charge estimator.
\end{enumerate}
%

\subsection{Control variables}
\label{deciders}
The following variables are used as inputs to the neural net in
order to provide it with some topological and \mbox{$b$-hadron}
specific information:
\begin{enumerate}
\item {$|\cos \theta|$} is included since it is correlated with the
quality of the secondary vertex reconstruction and hence with the
tagging power of the estimators relying on the separation between
tracks from the primary and secondary vertex.
\item {\em The reconstructed \mbox{$b$-hadron} momentum} is included
because
the relative accuracy of the various charge estimators depends
on the $b$-hadron momentum. An estimator of this momentum
is constructed from the jet closest to the line-of-flight of
the $b$-hadron.
The estimator
is the sum of the charged track
momenta in the jet, weighted by the probability that they come
from the secondary vertex, and the projections of the neutral momenta
onto the line-of-flight of the $b$-hadron. The missing energy
in the hemisphere
is also added.
\item{\em The reconstructed proper time of the \mbox{$b$-hadron}}~is
used based on the reconstructed \mbox{$b$-hadron} momentum and the
measured decay length.  The intention here is to incorporate the
increased probability of $B^0_{d}$ mixing at long proper times.
\item{\em The spread of track separation weights} i.e. the width
of the $w_i$ factors distribution in a given hemisphere.
This
allows the net to de-weight those charge estimators
which suffer from an ambiguous allocation of tracks to the primary and
secondary vertices in cases of high charged multiplicities and/or poor
vertexing.
\end{enumerate}

\begin{figure}[htbp]
\begin{center}\mbox{\epsfig{file=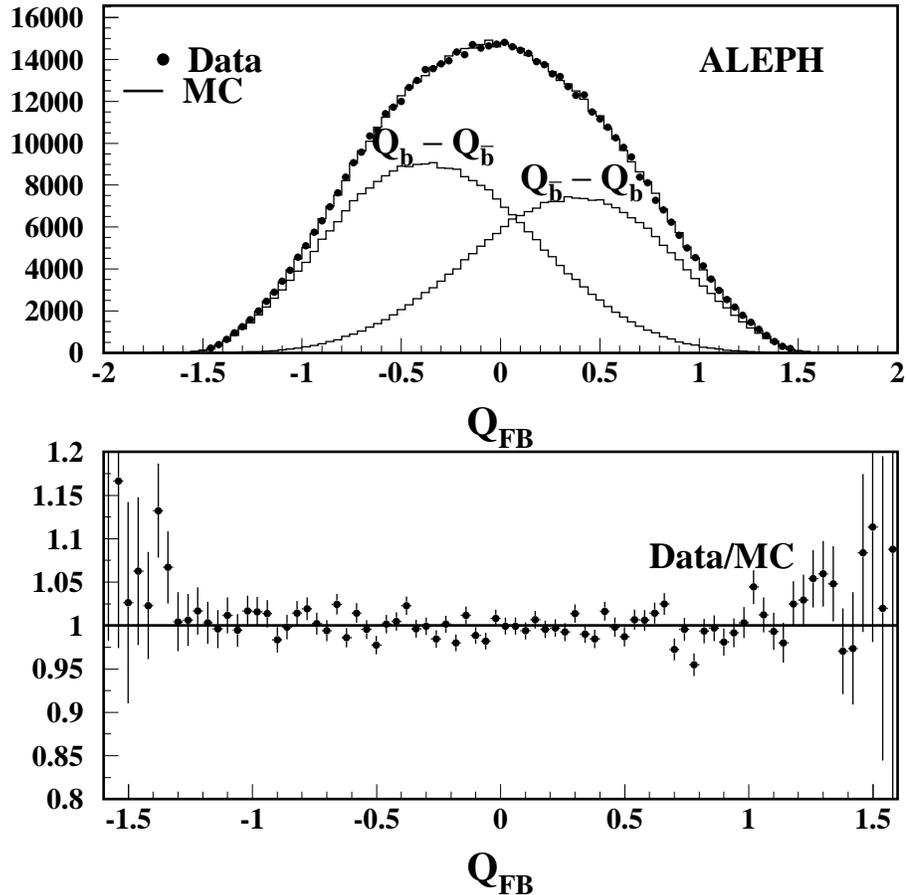,width=0.8 \textwidth}}\end{center}
\vspace{0.2 cm}
\caption{The charge difference between forward and backward hemispheres
measured in the selected event sample
labelled {\bf E} in Figure \ref{fig_b1b2}. The MC distribution corresponds
to an asymmetry of $A^b_{\rm FB} = 0.0967$.
\label{fig_qfb}}
\end{figure}

For the asymmetry calculation, the difference between the neural net
outputs of the two hemispheres
is used. This is shown in Figure~\ref{fig_qfb} together with the
MC prediction. The charge separation is seen to be reasonably well
simulated, but with a slightly larger width in data than in MC
(there is also a small difference in the average value of the
two distributions, but this is too small to be visible on the plot).
%

\section{Event selection}
\label{Ev sel}
The data set used for this analysis consists of four
million hadronic $Z$ decays recorded by ALEPH during the period 1991
to 1995 in a centre-of-mass energy range of $M_Z \pm 3 \,$GeV.
Events are selected according to the standard ALEPH hadronic event
selection  based on charged track information \cite{zpaper}.
This selection has an efficiency of $
97.5 \, \%$ and the backgrounds
from $Z$ decays to $\tau^+\tau^-$ and $\gamma \gamma$ interactions are
estimated to be $\sim 0.3$\% each.  These
backgrounds are reduced to the 10$^{-5}$ level after the application of
the subsequent cuts and can be safely neglected.

\begin{figure}[htbp]
\begin{center}\mbox{\epsfig{file=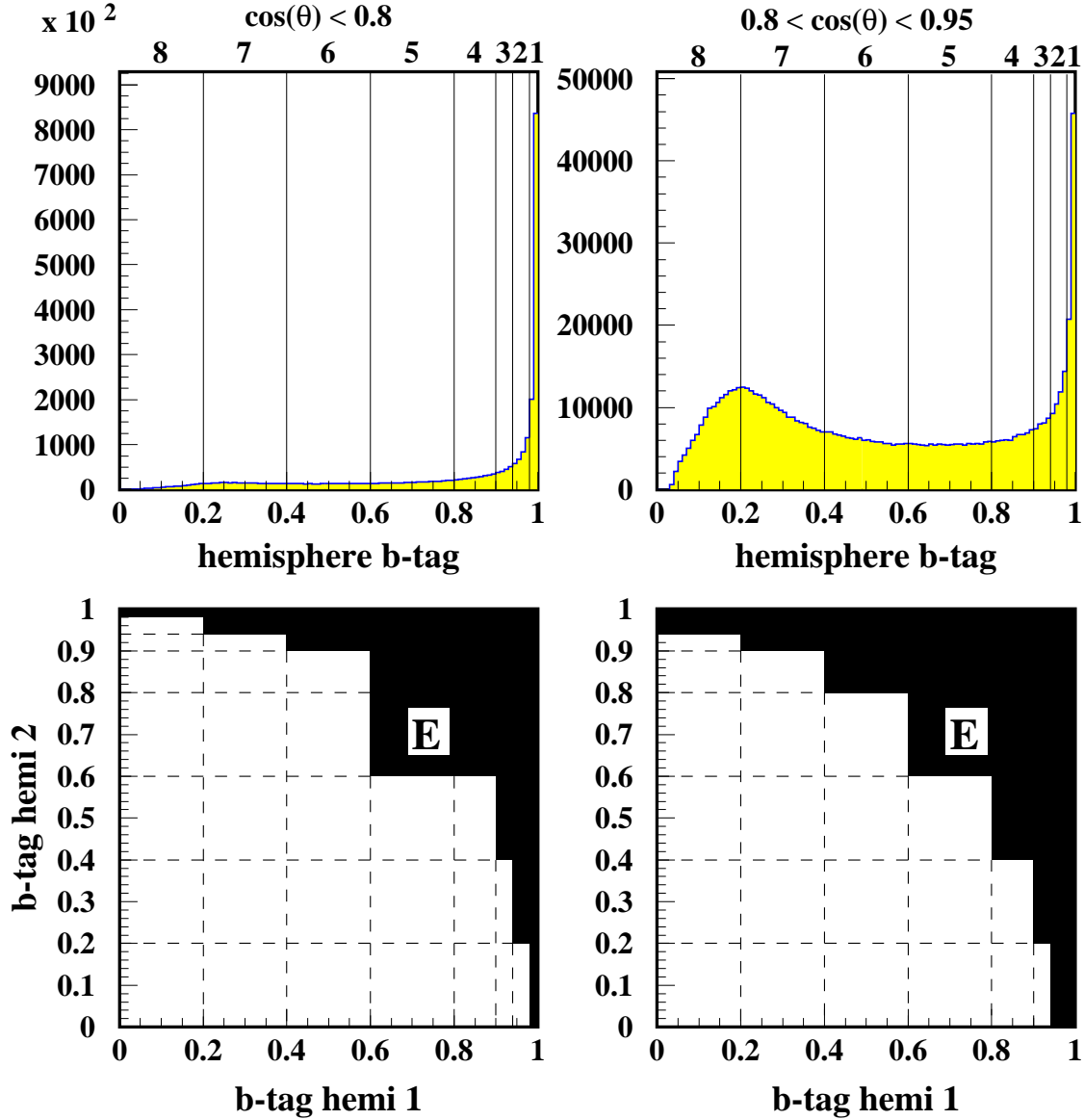,width=\textwidth}}\end{center}
\vspace{0.2 cm}
\caption{{\em Upper plots:}
Subdivision into $b$-tag intervals with numbers above the plots
enumerating the intervals. The shaded histogram is the expected
$b$-tag distribution for $b$-hemispheres, shown
separately for two regions of polar angle.
{\em Lower plots:} The event acceptance, shown shaded and
labelled {\bf E}, in the plane of the two hemisphere $b$-tags.
\label{fig_b1b2}}
\end{figure}

The average beam-spot position is determined every 75 events
and used
to constrain the \mbox{event-by-event} interaction point. In order to
remove lifetime information from the tracks, these are
projected onto the plane perpendicular to their parent jets (unless
they point behind the interaction point found in a first iteration).
Combining
these projections with the beam-spot position fixes the interaction
point to a precision of $50 \times 10 \times 60 \, \mu {\rm m}^3$ in
horizontal, vertical and beam directions respectively.

The thrust axis is determined in each event from all
charged tracks and calorimeter clusters using the ALEPH energy flow
algorithm \cite{ALEPH perf}.  To ensure that a good fraction of the event
is contained within the detector acceptance, the cosine
of the thrust axis polar angle
is required to be less than $0.95$.
In order to be able to evaluate the hemisphere $b$-tag and charge tag
it is required
that each hemisphere has
at least one charged track and at least one jet with
energy greater than 10~GeV. With these selection cuts a sample
of 3,734,425 hadronic $Z$ decays is retained
for the remainder of the analysis.

In order to make use of double-tag methods for determining flavour purities,
the $b$-tag range
between zero and one is divided into eight sub-intervals as shown in
Figure \ref{fig_b1b2}. The 672,111 events which are
located in the region indicated by the black area
in Figure \ref{fig_b1b2}, labelled {\bf E}, are the ones
accepted for
the $A_{\mathrm{FB}}$ analysis. The acceptance is less strict
in the forward region which is not covered by the vertex detector
and therefore has a broader distribution of $b$-tags.

\section{Determination of flavour purities}
\label{purfit}

The fractions of primary quark flavours in the selected event
sample are
measured in the data using the double
tag method, that has been used previously in e.g.\ the
ALEPH $R_b$ measurement \cite{rblife,rb5tags}.

In each of nine angular intervals, eight $b$-tag sub-intervals
are defined as indicated in
Figure \ref{fig_b1b2}, out of which only seven are mutually
independent and the eighth is considered as ``not tagged''.
These give rise to 28 doubly tagged event classes
and 7 singly tagged classes: $n_{ij}$ and $n_{i8}$, which are
related to the tagging efficiencies by:
\begin{eqnarray}
\nonumber
n_{ij} & = & \sum_{f} (2-\delta_{ij}) \: \epsilon^{f}_{i} \:
 \epsilon^{f}_{j}
 \: c^f_{ij} \: R_f \: n_{{\mathrm had}} \, ; \; \; \; \; \, i \leq j \, , \\
n_{i8} & = & \sum_{f} 2 \: \epsilon^{f}_{i} \: R_f
( 1 - \sum_{j \leq 7} \epsilon^{f}_{j} \, c^f_{ij} ) \: n_{{\mathrm had}}
 \, ; \; \; i \leq 7 \, , 
\label{eq_double_tag}
\end{eqnarray}
where $\epsilon^{f}_{i}$ is the probability
for a primary quark of flavour $f$ to be
located in interval $i$, $c^f_{ij}$ describes correlations
between the $b$-tag values in the two hemispheres,
$R_f$ is the $Z \rightarrow f \bar{f}$ fraction of the $Z$
hadronic decays and
$n_{\mathrm{had}}$ is the total number of events in the
sample under consideration.
The flavour index, $f$, runs over three flavour classes only
($uds$, $c$ and $b$), hence efficiencies and correlation factors
are averaged over the three light flavours.

Various assumptions are made to reduce the number of unknown
parameters in Equation~(\ref{eq_double_tag}). The hemisphere
correlations are taken from MC simulation and the values of
$R_f$ are taken from previous measurements.
Furthermore, it
is necessary to fix the
ratio of the two smallest efficiencies, where
$\epsilon^{uds}_{1} \approx 0.0008$ and
$ \epsilon^{uds}_{2} \approx 0.0013$
according to the simulation.
The {\em ratio} between these numbers is
taken from simulation.
\begin{table}[htbp]
\begin{center}
\begin{tabular}{|c|c|c|c|c|c|c|} \hline
  &
\multicolumn{3}{|c|}{Fitted purities} &
\multicolumn{3}{|c|}{Fit$-$MC} \\
$\cos (\theta)$ &   $uds$ & $c$ & $b$ & $uds$ & $c$ & $b$ \\
\hline 
$0.0-0.1$ & $157 \pm 46 \pm 3$  &
$ 1237 \pm 69 \pm 7$  &
$ 8606 \pm 34 \pm 15$  &
$-33$ & 96 & $-63$ \\
$0.1-0.2$ & $239 \pm 56 \pm 5$  &
$ 1170 \pm 77 \pm 7$  &
$ 8591 \pm 32 \pm 14$  &
 43 & 6 & $ -49$ \\
$0.2-0.3$ & $208 \pm 48 \pm 4$  &
$ 1156 \pm 69 \pm 7$  &
$ 8636 \pm 31 \pm 14$  &
 15 & 13 & $-28$  \\
$0.3-0.4$ & $246 \pm 55 \pm 5$  &
$ 1098 \pm 74 \pm 6$  &
$ 8657 \pm 31 \pm 14$  &
 64 & $-26$ & $-38$  \\
$0.4-0.5$ & $200 \pm 36 \pm 4$  &
$ 1064 \pm 56 \pm 6$  &
$ 8736 \pm 40 \pm 14$  &
 32 & 8 & $-40$ \\
$0.5-0.6$ & $147 \pm 42 \pm 3$  &
$ 992 \pm 63 \pm 6$  &
$ 8863 \pm 31 \pm 14$  &
 $-15$ & 22 & $-7$  \\
$0.6-0.7$ & $197 \pm 52 \pm 5$  &
$809 \pm 73 \pm 5$  &
$ 8994 \pm 29 \pm 14$  &
 42 & 14 & $-56$  \\
$0.7-0.8$ & $422 \pm 94 \pm 16$  &
$366 \pm 109 \pm 3$  &
$ 9212 \pm 30 \pm 14$  &
 238 & $-242$ & $-4$ \\
$0.8-.95$ & $405 \pm 203 \pm 5$  &
$883 \pm 250 \pm 8$  &
$ 8721 \pm 57 \pm 14$  &
 27 & 342 & $-369$  \\
\hline
\end{tabular}
\caption{Flavour purities of the event sample
labelled {\bf E} in Figure \ref{fig_b1b2} as obtained from the double tag fit.
The values are given in units of 10$^{-4}$. 
The second error is from MC statistics
contributing through the assumed hemisphere tag correlations.}
\label{tab_pur}
\end{center}
\end{table}

The remaining 20 unknown efficiencies $\epsilon^{f}_{i}$ are
determined from a fit using
Equations~(\ref{eq_double_tag})
leaving 15 remaining degrees of freedom in the fit.

The fitted efficiencies are combined to give event tag
efficiencies by summing over the fiducial region labelled {\bf E}
in Figure \ref{fig_b1b2}:
\begin{equation}
\epsilon_{E}^{f} \: = \: \sum_{i,j \in E} \epsilon^{f}_{i} \, \epsilon^{f}_{j}
 \, c^f_{ij} \, ,
\end{equation}
and from these efficiencies the three purities are constructed
as:
\begin{equation}
 {\cal P}_{E}^{f} \: = \: \frac{\epsilon_{E}^{f} R_f}{ \epsilon_{E}^{b} R_b +
 \epsilon_{E}^{c} R_c
 + \epsilon_{E}^{uds} (1- R_b - R_c) } \, ,
\label{eq_pevt}
\end{equation}
where the used $R_f$ fractions are shown in Table \ref{tab_smpar}.
The result of the fit
is shown in Table \ref{tab_pur}.
The covariance matrix of the fitted efficiencies is propagated to the
event-tag purities with
diagonal elements
also shown in
Table \ref{tab_pur} and correlation coefficients being typically $-90$\% ,
 +45\% and $-80$\% for the $uds$-$c$, the $uds$-$b$ and the $c$-$b$
purity correlations, respectively. The
$\chi^2$ of the nine independent fits at different $\cos\theta$
averages 0.99 
per degree of freedom.

Already from Figure \ref{fig_btag} it is seen that the
efficiencies in the highest $b$-tag bins are higher in the data than
in the MC. The fit determines that this enhancement is slightly
larger for light and charm quarks than for $b$-quarks,
resulting in a $b$ purity which is about 0.5\% lower than
predicted by the MC in the selected sample.

\section{Determination of quark charge separations}
\label{dbdc}

The mean values of the hemisphere charge
separations, $\delta_{f}$, are determined from
data in each bin of
$\cos \, \theta$ using the following procedure.

  The data sample is subdivided into
  14 bins with flavour compositions ranging from almost pure $uds$ to
  almost pure $b$ flavour. The
  sum of squares of the two hemisphere $b$-tags is used to define the
  bins and the flavour compositions of these new bins are derived from
  the fit to the data described in the previous section.

  In each bin of this event $b$-tag variable,
  $\bar{\delta}$ is measured according
  to Equation~(\ref{equation-dbar}).
  A fit is performed using the 14 $\bar{\delta}$ measurements
  and the right side of Equation~(\ref{equation-dbar}) with
  the flavour
  specific values $\bar{\delta}_{b}$, 
  $\bar{\delta}_{c}$ and $\bar{\delta}_{uds}$
  as free parameters. Since the $b$-tags in the 14
  bins bias the charge estimator differently, additional assumptions
  taken from MC are needed in order to compare the
  14 $\bar{\delta}$ measurements. For each flavour,
  the differences
  between the $\bar{\delta}$ in a given bin and the $\bar{\delta}$
  in the {\bf E}-region,
  $\beta = \bar{\delta}_{\mathrm{E}} - \bar{\delta}_{\mathrm{bin}}$,
  are held fixed at their MC values. Thus,
  for each flavour, the fit shifts $\bar{\delta}$ by a fixed amount
  in all bins, minimising the following $\chi^2$:
\begin{eqnarray}
  \chi^2 & = & \sum_{i=1}^{14} \frac{ \left(
 \bar{\delta}^2_{{\rm meas},i} -  \sum_{f=1}^{3} {\cal P}^f_i \left(
            \bar{\delta}_{f} \, r_f - \beta^f_i \right)^2 \right)^2 }
  { \left( \Delta \bar{\delta}^2_{{\rm meas},i} \right)^2 } \, ,
\label{eq_dbdc}
\end{eqnarray}
  where $i$ labels the $b$-tag bins and $f$ the flavours ($b$,
  $c$ and $uds$). The $\bar{\delta}_{f}$ appearing in the $\chi^2$
  refers to MC events in the {\bf E}-region and
  the correction factors $r_f$
  are the three free
  parameters of the fit.

\begin{figure}[htbp]
\begin{center}\mbox{\epsfig{file=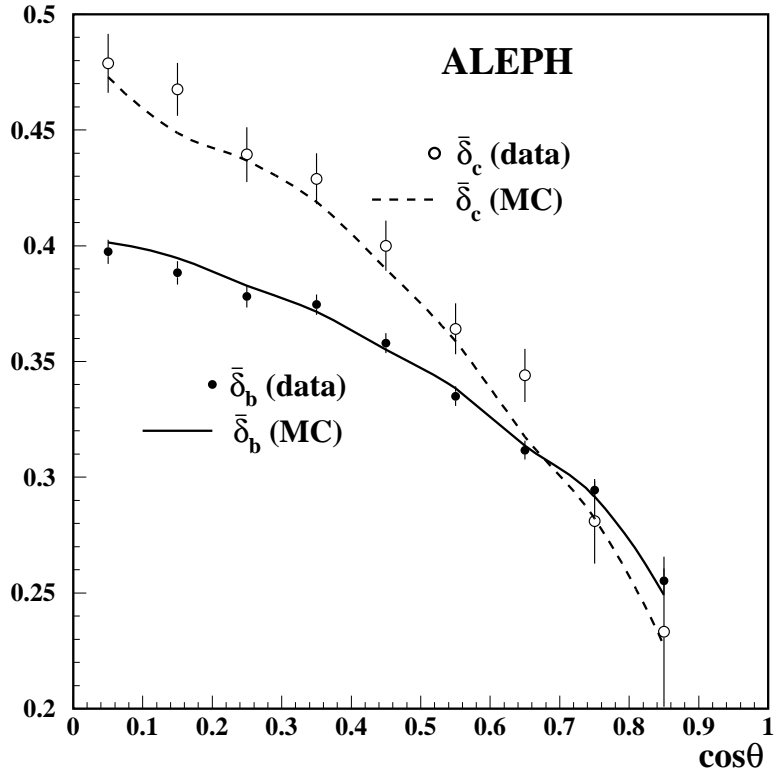,width=0.7 \textwidth}}\end{center}
\vspace{0.2 cm}
\caption{The fitted values of $\bar{\delta}_{b}$
and $\bar{\delta}_{c}$ as a function of polar angle together with the
values predicted by the MC.
\label{fig_dbar}}
\end{figure}

  The inputs to the fit, averaged over all $\cos \theta$,
  are shown in Table~\ref{table-purities} for $b$- and $c$-quarks
  together with
  the flavour combined output and the MC
  prediction. The fit reproduces accurately the $b$-tag dependence
  of  $\bar{\delta}$. The angular dependence
  of the fitted parameters is shown in
  Figure~\ref{fig_dbar} and, since the ratios between fitted and
  predicted values are consistent with being constant over angles, their
  average values are quoted below:
\begin{eqnarray}
r_{b} = 0.995 \pm 0.004 \nonumber \, ,\\
r_{c} = 1.031 \pm 0.011 \nonumber \, , \\
r_{uds} = 1.009 \pm 0.009 \nonumber \, .
\end{eqnarray}

  \begin{table}[htbp]
  \begin{center}
  \begin{tabular}{|c|c|c|c|c|c|c|} \hline
${\cal P}_{c}$ &
${\cal P}_{b}$ &
 $\beta_c$ & $\beta_b$ &
{\em Observed   } & 
{\em Expected } & \em Fitted \\ 
 & & & & $\bar{\delta}$    &  $\bar{\delta}$ &
       $\bar{\delta}$  \\ \hline
5.2& 0.3& 0.0676& 0.1650& $0.4413 \pm 0.0061$& 0.4407& 0.4430 \\
8.6& 0.8& 0.0808& 0.1559& $0.3877 \pm 0.0039$& 0.3851& 0.3878 \\
11.8& 1.6& 0.1016& 0.1663& $0.3347 \pm 0.0036$& 0.3341& 0.3371 \\
15.5& 3.0& 0.1119& 0.1493& $0.3115 \pm 0.0035$& 0.3083& 0.3124 \\
20.0& 5.0& 0.0979& 0.1496& $0.3039 \pm 0.0033$& 0.3003& 0.3042 \\
24.9& 8.2& 0.0884& 0.1284& $0.3030 \pm 0.0031$& 0.2960& 0.3001 \\
29.7& 12.2& 0.0729& 0.1186& $0.3048 \pm 0.0030$& 0.2982& 0.3029 \\
34.4& 18.4& 0.0572& 0.1118& $0.3011 \pm 0.0028$& 0.2994& 0.3026 \\
38.4& 27.7& 0.0465& 0.0945& $0.3080 \pm 0.0026$& 0.3028& 0.3072 \\
36.2& 45.3& 0.0237& 0.0813& $0.3144 \pm 0.0023$& 0.3092& 0.3130 \\
17.4& 77.7& 0.0098& 0.0445& $0.3113 \pm 0.0021$& 0.3146& 0.3157 \\
12.5& 85.8& $-$0.0104& 0.0290& $0.3276 \pm 0.0028$& 0.3272& 0.3275 \\
7.0& 92.5& $-$0.0252& 0.0091& $0.3411 \pm 0.0029$& 0.3421& 0.3420 \\
1.2& 98.8& $-$0.0802& $-$0.0455& $0.3910 \pm 0.0019$& 0.3941& 0.3886 \\
 \hline
\end{tabular}
\caption{From left to right:
 bin-by-bin flavour purities
 (${\cal P}$ in percent), biases
  ($\beta = \bar{\delta}_{\mathrm{E}} - \bar{\delta}_{\mathrm{bin}}$),
 measured, MC expected and
 fitted values of $\bar{\delta}_{\mathrm{bin}}$.
  The numbers in this
  table are integrated over all polar angles.}
  \label{table-purities}
\end{center}
\end{table}

  The average correlation coefficients
  of the error matrix are $-$64\% between $uds$ and $c$,
  25\% between $uds$ and $b$ and $-$49\% between $c$ and $b$.
 The average $\chi^2$ probability
  is 0.72 over the independent fits in nine bins of polar angle.

  The fitted values of $\bar{\delta}_{f}$ are translated into
  the charge
  separations, $\delta_{f}$ by
  Equation~(\ref{equation-db from dbarb}), where the
  charge correlation correction factors, $(1+k_{f} )$,
  are taken from MC. These factors vary
  with polar angle, averaging
  1.086 for $c$-quarks
  and 1.092 for $b$-quarks.  

  In the fit of
  $A_{\mathrm FB}^b$ from Equation~(\ref{eq_afb}), the
  values for the charge separations,
  $\delta_{f}(\theta )$, are thus the MC values corrected
  by the $r_{f}(\theta )$ factors.
  The ratios between individual light
  quark charge
  separations are taken from MC. In summary, the following
  values are used (averaging over polar angle and showing only
  the diagonal elements of the statistical error matrix):

\begin{eqnarray}
 \delta_{b} & \: = \: & -0.3143 \pm 0.0014 \nonumber \, , \\
 \delta_{c} & \: = \: & +0.3733 \pm 0.0039 \nonumber \, , \\
 \delta_{s} & \: = \: & -0.1929 \pm 0.0016 \nonumber \, , \\
 \delta_{d} & \: = \: & -0.1547 \pm 0.0013 \nonumber \, , \\
 \delta_{u} & \: = \: & +0.2598 \pm 0.0022 \nonumber \, . \\
 \nonumber
\end{eqnarray}
 This charge separation corresponds to an average event mis-tag
 of 25.6\% for $b\bar{b}$ events. The result of the asymmetry
 fit to the peak data is
\begin{eqnarray}
\nonumber                      
A^b_{\mathrm{FB}}(\sqrt{s} = 91.232~{\rm GeV})
 \: = \: 0.0997 \pm 0.0027(stat).
\end{eqnarray}

\section{Systematic errors}
\label{syserrors}

  The determinations of the sample purity, the charge separation
  and the forward--backward asymmetry itself all rely to some extent
  on samples of Monte Carlo events. The features of the simulation
  most relevant for the analysis have been identified and varied
  within their errors in order to estimate the
  systematic uncertainty of the measurement as described in detail in
  the following sections.

\boldmath
\subsection{The uds purity at large $b$-tags}
\unboldmath
 In the fit yielding the flavour purities of the selected event sample
 it was assumed that ratio of the tiny efficiencies
 for selecting a light flavour
 with $b$-tag $> 0.98$ and for
 selecting a light flavour
 with $0.94 < b$-tag $ < 0.98$  was correctly predicted by the MC
 to be 0.62, averaged over polar angle. A systematic uncertainty on this
 ratio of $\pm 30$\% is estimated by varying the track parameter
 smearing and the gluon splitting rates to heavy flavours in the
 MC. Propagating this uncertainty to the asymmetry
 yields a variation of $\pm 0.00037$.

\boldmath
\subsection{Correlations between hemisphere $b$-tags}
\unboldmath
The correlation factors ($c^f_{ij}$ in Equation~(\ref{eq_double_tag}))
are taken from Monte Carlo simulation.
The influence of the correlations on the purities, via the
fitted flavour efficiencies, is limited by Equation~(\ref{eq_pevt})
because the $b$-quark purity is high, 88\% on the average. The correlation
coefficient with the largest impact on the $b$ purity is the negative
correlation for observing both hemispheres in the highest $b$-tag
bin: $c^b_{11}-1 = -0.02$ in the central region of the detector. It
is due to the pull on the shared reconstructed primary vertex exerted by a
secondary vertex from $b$-hadron decay.
If that coefficient is changed by an amount
$\pm 0.01$, the $b$ purity changes by the amount $\pm 2.3 \cdot 10^{-4}$
in the central region.
The other
coefficients carry impacts on the flavour purities of size $10^{-4}$ or less
for a one percent change in the coefficient.

In order to estimate the uncertainty in the correlation
coefficients, the input parameters to the MC given
in Table~\ref{table-physpar}
are
changed, one by one by the re-weighting technique,
to the one $\sigma$ higher values.
Two additional re-weighted MC samples are studied in order
to cover effects of QCD gluon radiation and of fragmentation. In
one sample
the angles separating the two selected $b$-jets are
forced to agree with data and in another sample the charged
multiplicity of the primary vertex is forced to agree with data.
The largest
effects on the $b$ purities arise from changing
the  $x_b$ beam energy fraction,
the charge
particle multiplicities
and the inter-jet angles, ranging from typically $\pm 10^{-5}$ in
the central region to $\pm 10^{-4}$ at the most forward angles.
The individual changes
are added in quadrature to produce the total variation in
the asymmetry of 0.00001.

There could be effects other than $b$-physics parameters causing
an inaccurate simulation of the correlation factors. The possibilities
include inaccuracies in the simulation of vertexing and of
QCD effects.
Therefore, a study of additional systematic errors is carried out
along the lines of the ALEPH $R_b$ measurement \cite{rblife,rb5tags}.
Here, the contribution to the correlation from a given variable
$v$ is calculated as:
\begin{equation}
c_{ij}(v) = \frac{\int (f_i (v)*g_j (v) + g_i (v)*f_j (v)) dv}
 {2 \ (\int f_i (v)dv) \ (\int f_j (v)dv) }
\label{eq_corvar}
\end{equation}
where $f_i (v)$ is the fraction of hemispheres, among all hemispheres
with the variable situated between $v$ and $v+dv$, having the tag $i$.
Similarly $g_i (v)$ is the corresponding fraction with the opposite
hemisphere tagged by tag $i$. The chosen variables, $v$, are the
thrust polar and azimuthal angles, an estimator of the  $b$-hadron
momentum and the six input variables to the $b$-tagging neural network.
Since the most important correlations are those involving $b\bar{b}$
events, the ten most significant of those are studied.
In order to avoid having to subtract a very large
non-$b$ background in some bins,
a soft $b$-tag is placed before evaluating the integrals.
This soft $b$-tag leaves a 64\% pure $b$-sample.

For the azimuthal angle,
where no significant correlation is expected,
the result of Equation~(\ref{eq_corvar})
is indeed unity within a precision of $10^{-3}$. The rest of the
variables do produce correlations, but the differences between
data and MC are all between $10^{-3}$ and $10^{-2}$ and hence 
their contribution to the error on the measured asymmetry
(via the flavour purities)
is again very small, in total 0.00001.

Until now only correlations in $b\bar{b}$
events have been studied.
Although the individual correlations
for the lighter quarks have very little impact, the possibility
of a combined effect of a large number of correlations has been
investigated by simply ignoring all of them in the fit (except at the
very most forward angle where geometrical effects are important).
This results in a change in $A^b_{\mathrm{FB}}$ of +0.00002.
A summary of all errors
connected with the estimation of flavour purities is
given in Table~\ref{tab_purerr}.

\begin{table}[htbp]
\begin{center}
\begin{tabular}{|c|c|} \hline
Statistical error in the purity determination& $\pm 0.00039$  \\
Light quark $b$-tag efficiencies & $\pm 0.00037$  \\
Hemisphere $b$-tag correlations & $\pm 0.00003$  \\
\hline
Total systematic error due to purities    & $\pm 0.00054$  \\        
\hline                                                                       

\end{tabular}
\caption{ Summary of systematic errors
on the  $A^b_{\rm FB}$ measurement from the determination of flavour
purities.}
\label{tab_purerr}
\end{center}
\end{table}

%
\subsection{Systematic uncertainties on charge separations}
\label{dbdc_errors}
Various assumptions contribute to systematic uncertainties on
the determination of the charge separations, $\delta_f$.
In a first step,
$\bar{\delta}_b$, $\bar{\delta}_{c}$ and 
$\bar{\delta}_{uds}$ are
extracted from a fit to data, where the assumed flavour purities and
charge bias corrections in the different $b$-tag bins contribute to
the systematic error together with the statistical errors of the fit.
In a second
step, the constant of proportionality, $( 1 + k_f )$,
between $\bar{\delta}_f$ and
$\delta_f$ contributes to the uncertainty. These systematic uncertainties are
summarised in Table \ref{table-dbdcsystsum} and discussed in detail
in the following.

\begin{itemize}
\item{\em Flavour purities}~-~ This is evaluated by using the purities
predicted by the MC instead of the fitted purities. The
full changes in the fitted values of
$\bar{\delta}_f$ are propagated to the error on the asymmetry.  
\item{\em Corrections for $b$-tag bias}~-~
  Two overlapping procedures are used to estimate this uncertainty:
\begin{enumerate}
\item Each of the $b$-physics parameters listed in
  Table~\ref{table-physpar} are increased one at the time by one
  standard deviation, using re-weighted versions of the MC, and
  the differences in the {\em fitted} $\delta_{b}$ and
  $\delta_{c}$
  are propagated to the asymmetry and entered as systematic errors in Table
    \ref{table-dbdcsystsum} . These estimates include the effect
  of the parameters on the charge correlation corrections
  discussed below. As in the case of hemisphere tag correlations,
  the effect of re-weighting the MC to agree with the observed primary
  vertex multiplicity and the observed distribution of inter-jet angles
  are also included. Finally, the $Vector / (Vector + Pseudoscalar)$
  ratio in $D$-meson production is varied within its experimental error
  of 0.05 \cite{charm}.

\item The values of $\bar{\delta}_{c}$ and $\bar{\delta}_{uds}$
  are fixed at their MC values instead of being left free in the fit.
  In this case Equation~(\ref{equation-dbar}) can be solved
  directly for $\bar{\delta}_{b}$ in the {\bf E}-region
  without any assumptions concerning bias corrections.
  The full difference between the result of this and the default result
  is taken as an independent
  contribution to the systematic error. This
  corresponds to an excursion in the fitted
  value of $\bar{\delta}_{c}$ by almost two times the other uncertainties
  on this quantity.

\end{enumerate}
\item{\em Charge correlation corrections}~-~
 One contribution to the systematic error from the $(1+k_f)$
 factors is found by varying
 the input parameters of the MC as described above.

 Another contribution
 is estimated, like in the case of hemisphere $b$-tag correlations,
 from the difference between data and MC
 when building on the {\bf E}-sample a
 ``projection'' of the hemisphere
 charge correlations
 along a series of observables, as follows:

\begin{equation}
 (1+k_{v})^2 = \frac{\int \left( Q_{\rm same}^{+} (v)*Q_{\rm opp}^{-} (v) +
                Q_{\rm same}^{-} (v)*Q_{\rm opp}^{+} (v) \right) P (v) dv}
              {2 \langle Q^{+} \rangle \langle Q^{-} \rangle } \, ,
\label{eq_chcor}
\end{equation}

  where $ Q^{\pm} (v)$ is the average positive or negative hemisphere
  charge, given the variable $v$ in either the same or the
  opposite hemisphere, and $P(v)$ is the probability density of $v$.
  Eight variables are chosen that are weakly correlated
  while covering the variables relevant for the charge determination.
  These are the thrust, the thrust axis direction,
  some inputs to the charge tagging
  neural net, the $b$-tag and the $b$-hadron momentum.
  The differences in $(1+k_{v})$ between data and MC are 
  propagated to
  Table~\ref{table-dbdcsystsum}.

This procedure relates to the simulation of charge correlations
for events in the selected sample consisting mostly of $b\bar{b}$
events. Previous studies using the
jet-charge technique \cite{aleph qfb, original}
have assessed the uncertainty in $k_c$
separately. Using a weighting power of $\kappa = 0.5$ the value
$k_c = 0.085 \pm 0.022$ was found, where the error includes
systematic uncertainties. This value is equal to the charm charge
correlation in the present study and its error is propagated as
an additional contribution to Table~\ref{table-dbdcsystsum}.

\item{\em Light quark charge separations}~-~
The light quark charge separations, $\delta_{u}$,
$\delta_{d}$ and $\delta_{s}$, are taken from Monte
Carlo, but corrected by the overall correction factor from
the fit. 
Systematic uncertainties are ascribed according to the
results of~\cite{aleph qfb} which represent relative uncertainties of 
2.1\%, 4.0\% and 1.5\% for $u$, $d$ and $s$ respectively. These
errors are much larger than the error on the combined
light quark $\bar{\delta}_{uds}$ from the fit. However,
when propagated through to $A_{\rm FB}^b$, the systematic uncertainties
arising from light quark charges are small, in total 0.00003.

\item{\em Track parameter smearing}~-~
  The extra smearing and hit inefficiency mentioned in Section
 \ref{nnbtag} also affects the charge tag. The effect of dropping
 these corrections to the MC altogether
 results in a change in the asymmetry of +0.00016
 which is taken as an additional contribution to the systematic error.

\end{itemize}
  \begin{table}[htbp]
  \begin{center}
  \begin{tabular}{|l|c|} \hline
\rule{0mm}{5mm} Source of systematic uncertainty &

$\Delta A_{\rm FB}^b $ \\ \hline \hline
\rule{0mm}{4mm} Statistics in the $\bar{\delta}_{f}$
 determination  & 0.00074 \\
\hline
flavour purities                         & 0.00015 \\
\hline
$< x_{b} >$ beam energy fraction & 0.00007 \\
$< x_{c} >$ beam energy fraction & 0.00000 \\
$n_{ch}$ in $b$-hadron decay & 0.00003 \\
$B_{s}$ fraction                   & 0.00004 \\
$\Lambda_{b}$ fraction             & 0.00004 \\
$B^+$ lifetime                         & 0.00003 \\
$B^0$ lifetime                         & 0.00002 \\
$B_{s}$ lifetime                   & 0.00002 \\
$b$-baryon lifetime             & 0.00003 \\
$g \rightarrow {b}\bar{b}$ rate & 0.00008 \\
$g \rightarrow {c}\bar{c}$ rate & 0.00004 \\
primary multiplicity                   & 0.00001 \\
inter-jet angles                       & 0.00004 \\
$\frac{V}{V+P}$ for charm              & 0.00019 \\
\hline
fixed $\delta_{c}$ and $\delta_{uds}$
                                             & 0.00037 \\
\hline
charge correlations due to thrust       & 0.00001 \\
charge correlations due to $\cos\theta$ & 0.00008 \\
charge correlations due to $\phi$       & 0.00007 \\
charge correlations due to $b$-hadron momentum      & 0.00002 \\
charge correlations due to $b$-tag      & 0.00000 \\
charge correlations due to jet-charge   & 0.00012 \\
charge correlations due to primary vertex charge   & 0.00010 \\
charge correlations due to secondary vertex charge   & 0.00011 \\
charge correlations due to kaon charge   & 0.00000 \\
charm quark charge correlations          & 0.00016 \\
\hline
light quark charge separations          & 0.00003 \\
\hline
track parameter smearing                & 0.00016 \\
\hline
\hline
Total systematic error due to charge separations                                     & 0.00093 \\
\hline
\end{tabular}
  \caption{Summary of systematic error contributions
    to the  $A^b_{\rm FB}$ measurement
    from the determination of charge separations.}
  \label{table-dbdcsystsum}
\end{center}
\end{table}

\boldmath
\subsection{Experimental systematic errors on $Q_{\mathrm{FB}}$ }
\unboldmath

For the measuring apparatus itself to produce a fake
$Q_{\mathrm{FB}}$, a forward-backward asymmetric bias in
the charge measurement
is needed. Such an effect could come from an asymmetry in the
detector material, since nuclear interactions of the produced particles
with this material give rise to an excess of
charge which is measured by the total charge, $Q_{\rm tot}$,
in data \cite{aleph qfb}.
The asymmetry of the material is estimated
by measuring the forward-backward asymmetry of photons that have
converted to $e^+ e^-$ pairs in the detector.  This asymmetry
is multiplied by the material charge component
at each angle and subtracted from $Q_{\mathrm{FB}}$, causing
a shift in $A^b_{\mathrm{FB}}$ by $0.00016 \pm 0.00032$
which is taken as a contribution to the systematic error.

Another possible source of error is the modelling of the magnetic field
in the extreme forward region. This is known to cause a slight difference
in the momenta measured for positive and negative particles at momenta
close to the beam energy and a correction, which is not quite 
forward-backward symmetric, is applied as default on the data
to take out this effect. However, the effect on 
$A^b_{\mathrm{FB}}$ from dropping the correction is very small 
($0.00002$) and this is added to the error.

\subsection{Electroweak observables}
  Assumptions regarding electroweak observables
  other than
  $A^b_{\mathrm{FB}}$, e.g.\ $R_b$ and $A^c_{\mathrm{FB}}$, belong
  to a special class, since they are important for the Standard Model
  interpretation of the result. Their
  values used in this analysis are taken from
  the Standard Model and are listed in Table~\ref{tab_smpar}
  together with their impacts on the measured value of
  $A^b_{\mathrm{FB}}$.
  In case of the branching fractions $R_f$, the variation in
  one flavour is compensated by changes in the light quark $R_f$'s 
  ($f=uds$) according to their relative magnitudes.

\begin{table}[htbp]
\begin{center}
\begin{tabular}{|c|c|c|c|c|} \hline
\rule{0mm}{6mm} {\em Parameter} {\bf P} &
$\sqrt{s} = 89.416$ GeV & $\sqrt{s} = 91.232$ GeV  & $\sqrt{s} = 92.945$ GeV &
$\partial A^b_{\rm FB}/ \partial $ {\bf P} \\
\hline
$R_u$ &  & 0.17211 & & $-$     \\
$R_d$ &  & 0.22025 & & $-$     \\
$R_s$ &  & 0.22025 & & $-$     \\
$R_c$ &  & 0.17154 & & $+0.018$ \\
$R_b$ &  & 0.21585 & & $-0.440$ \\
\hline
$A^u_{\rm FB}$ & $-0.03342$ & 0.06507 & 0.12445 & $+0.006$ \\
$A^d_{\rm FB}$ & $0.06141$ & 0.09982 & 0.12447 & $-0.001$ \\
$A^s_{\rm FB}$ & $0.06150$ & 0.09983 & 0.12908 & $-0.001$ \\
$A^c_{\rm FB}$ & $-0.03350$ & 0.06513 & 0.12451 & $+0.103$ \\
\hline
\end{tabular}
\caption{ Summary of Standard Model inputs used in the measurement and 
their influence on the measured $A^b_{\rm FB}$.
\label{tab_smpar}}
\end{center}
\end{table}

\subsection{QCD correction}
The radiation of gluons from the original quark pair can in
principle lower the asymmetry.
The effect is expected to be small
with the present method, where the thrust-defined
hemisphere charges are
calibrated back to the charges of the primary quarks in
these hemispheres using partly data and partly MC.
However, this built-in correction may not be
sufficient, since the hemispheres defined by the
$q\bar{q}$ axis are not identical to those defined by the thrust.
Therefore the asymmetry measured by repeating
the analysis on the high statistics
$b\bar{b}$ MC is compared with the generator level asymmetry in the
same selected sample of MC events. The generator level asymmetry is
found to be higher by a factor
$1.0027 \pm 0.0009$.

The most accurate calculation available of the full QCD
correction in an unbiased sample of $b\bar{b}$ events results
in a correction factor
$1 + C_{\rm QCD} = 1.0354 \pm 0.0063$ \cite{qcd,qed}.
The ratio of $C_{\rm QCD}$ to the same
quantity calculated with the ALEPH MC is
used to scale the residual QCD
correction. Hereby the residual correction factor becomes
$1.0034 \pm 0.0019$, where the error takes into
account the statistical uncertainty and the fraction of the
theoretical error which affects this analysis.

\begin{table}[htbp]
\begin{center}
\begin{tabular}{|c|c|} \hline

Flavour purities & $\pm 0.00054$  \\
Charge separations & $\pm 0.00093$  \\
Detector asymmetries & $\pm 0.00016$  \\
QCD correction       & $\pm 0.00019$ \\
\hline
Total systematic error     & $\pm 0.00110$  \\        
\hline                                                                       
\end{tabular}
\caption{ Summary of systematic errors
on the  $A^b_{\rm FB}$ measurement.}
\label{tab_sys}
\end{center}
\end{table}

  A summary of all
  the systematic errors is given in Table~\ref{tab_sys}.

\section{Cross-checks}
\label{cross}
\boldmath
\subsection{Simultaneous fit to $R_b$}
\unboldmath
The selection efficiency fit which yields the
flavour purities of the selected event sample has been checked
by performing
the fit on MC and verifying that the input efficiencies are indeed
returned exactly. It has furthermore been checked that the distribution of
$b$-tags in the MC agrees with that measured in the data after re-weighting
the MC by the ratio of fitted to simulated flavour efficiencies.

Finally, the overall $b\bar{b}$
fraction, $R_b$, is left free in the fit. In order to get a stable
fit, it is necessary to tie down some of the previously free
parameters, and therefore the small $b$-hemisphere efficiencies for
$b$-tags less than 0.6 are taken from MC in this case. The result
(for the three million preselected hadronic $Z$ decays at peak energy) is
$R_b = 0.2174 \pm 0.0008 (stat)$.
This measured value is consistent with the world average ($0.21643 \pm 0.00073$
\cite{qed}).
The change in the asymmetry from letting $R_b$ float in the fit
is $-0.0005$, in agreement with the sensitivity shown in
Table~\ref{tab_smpar} and consistent with the
uncertainty of the fit.

\subsection{Lepton tagged events}

\begin{figure}[htbp]
\begin{center}\mbox{\epsfig{file=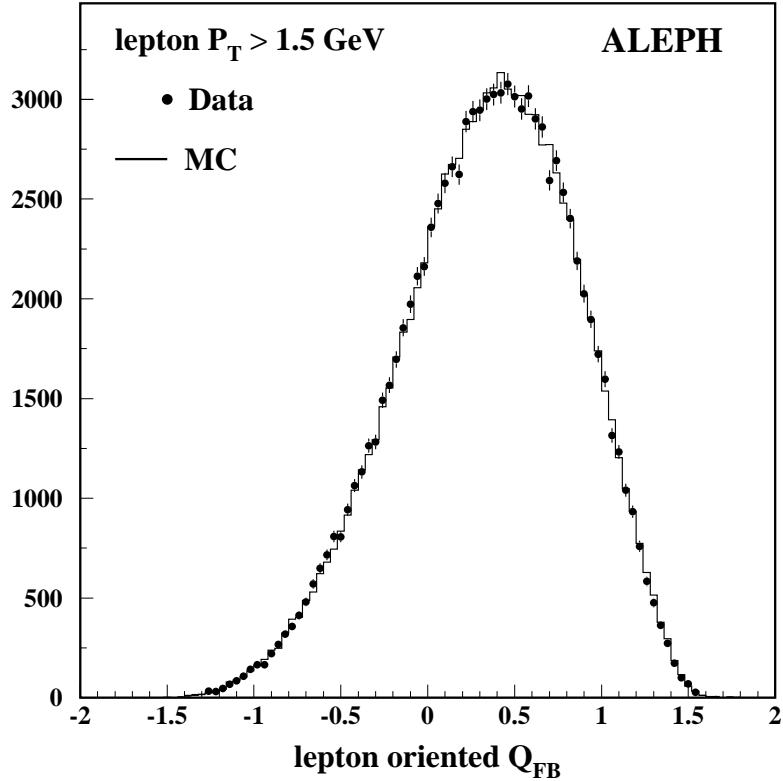,width=0.7 \textwidth}}\end{center}
\vspace{0.2 cm}
\caption{The difference between the two
hemisphere charges, signed by the lepton charge,
for events in the {\bf E} sample
containing a high-$p_T$ lepton.
\label{fig_lepton}}
\end{figure}

For the $\bar{\delta}$ fit, it has also been checked that it
returns the input  $\bar{\delta}_f$'s when applied to the MC.
 
The sample of events with an identified high--$p_T$ lepton
offers an additional check on the
assumptions entering the $b$-charge tag. This check is,
however, limited by uncertainties in the lepton tagging due to
$c \rightarrow l$, $b \rightarrow c \rightarrow l$ and
$B^{0} \rightarrow \bar{B}^{0}$ transitions as well as lepton
mis-identification.
These uncertainties are reduced to below the
$10^{-2}$ level
by placing a high--$p_{T}$ cut of 1.5 GeV/$c$ on
the lepton sample. The distribution of the signed difference between
the two hemisphere charges in such events is shown in Figure~\ref{fig_lepton}.
The sign of the lepton provides the
$b\bar{b}$ orientation in the event and allows a measurement
of $\delta_b = -0.3395 \pm 0.0016$, after subtracting the 3.8\%
non-$b$ background
in this sample. This is a factor $0.990 \pm 0.006$ smaller than the
same quantity measured in the $q\bar{q}$ MC. The corresponding ratio
between the $\bar{\delta}_b$'s was found to be  $0.995 \pm 0.004$
in Section
\ref{dbdc}. This then provides a check on the MC prediction of
$k_b$:
\begin{eqnarray}
\nonumber
\frac { (1 + k_b ) }{ (1 + k_b)^{\mathrm{MC} } } =
  \frac{ \delta_b  }{ \delta_b^{\mathrm{MC}} } /
  \frac{ \bar{\delta}_b  }{  \bar{\delta}_b^{\mathrm{MC}} }  =
0.995 \pm 0.007 \, (stat) \, .
\end{eqnarray}

\subsection{Stability checks}

In order to study the stability of the analysis the accepted region
in the plane of the two $b$-tags is varied over a wide range. The
results for peak data
shown in Table~\ref{vary_cuts} are consistent within
statistical errors.

\begin{table}[htbp]
\begin{center}
\begin{tabular}{|c|c|c|c|c|} \hline
\rule{0mm}{6mm} events selected & $b$ purity & raw $A^b_{\rm FB}$ & purity corrected &
 charge corrected \\
\hline
380086 & 0.967 & 0.0999 & 0.1006 & 0.1002 $\pm$ 0.0031 \\
491011 & 0.938 & 0.0987  & 0.0996 & 0.0994 $\pm$ 0.0028 \\
{\bf 596618} & {\bf 0.886} & {\bf 0.0990} &
{\bf 0.0999} & {\bf 0.0997} $\pm$ {\bf 0.0027} \\
643773 & 0.853 & 0.0995 & 0.1006 & 0.1007 $\pm$ 0.0027 \\
797132 & 0.749 & 0.1001 & 0.1006 & 0.1007
 $\pm$ 0.0027 \\
\hline
\end{tabular}
\caption{ Results on peak data for various event selections
characterised by the
sizes of the samples and their (predicted) $b$ purities.
The event selection shown
in bold-face is the chosen one, the {\bf E}-sample.
}
\label{vary_cuts}
\end{center}
\end{table}

The measurements performed for each LEP~I year in the data recorded
at $Z$ peak energies are
shown in Table \ref{vary_time}. They are seen to be
consistent with being constant.

\begin{table}[htbp]
\begin{center}
\begin{tabular}{|c|c|} \hline
\rule{0mm}{6mm} year &  $A^b_{\rm FB}$ \\
\hline
1991 & 0.0884 $\pm$ 0.0125 \\
1992 & 0.0992 $\pm$ 0.0063 \\
1993 & 0.1079 $\pm$ 0.0075 \\
1994 & 0.0997 $\pm$ 0.0040 \\
1995 & 0.0966 $\pm$ 0.0072 \\
\hline
average & 0.0997 $\pm$ 0.0027 \\
\hline
\end{tabular}
\caption{ Results on peak data for each LEP~I year.
}
\label{vary_time}
\end{center}
\end{table}

\section{Results and conclusions}
\label{results}

\begin{figure}[htbp]
\begin{center}\mbox{\epsfig{file=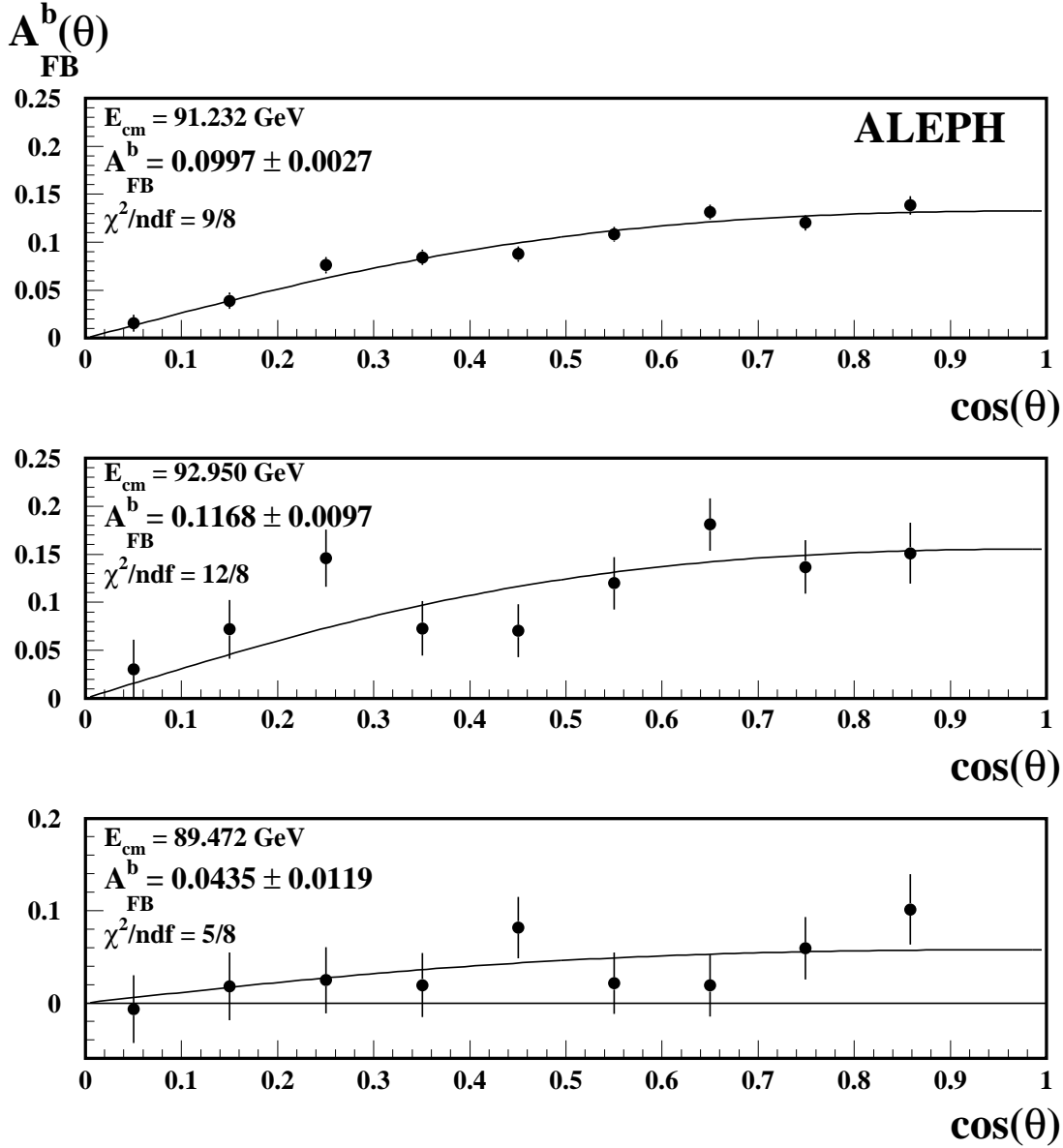,width=1.0 \textwidth}}\end{center}
\vspace{0.2 cm}
\caption{ The forward-backward $b\bar{b}$ asymmetry as a function of
thrust polar angle at three center-of-mass energies.
The errors shown are statistical only.}
\label{fig_costh}
\end{figure}

\begin{figure}[htbp]
\begin{center}\mbox{\epsfig{file=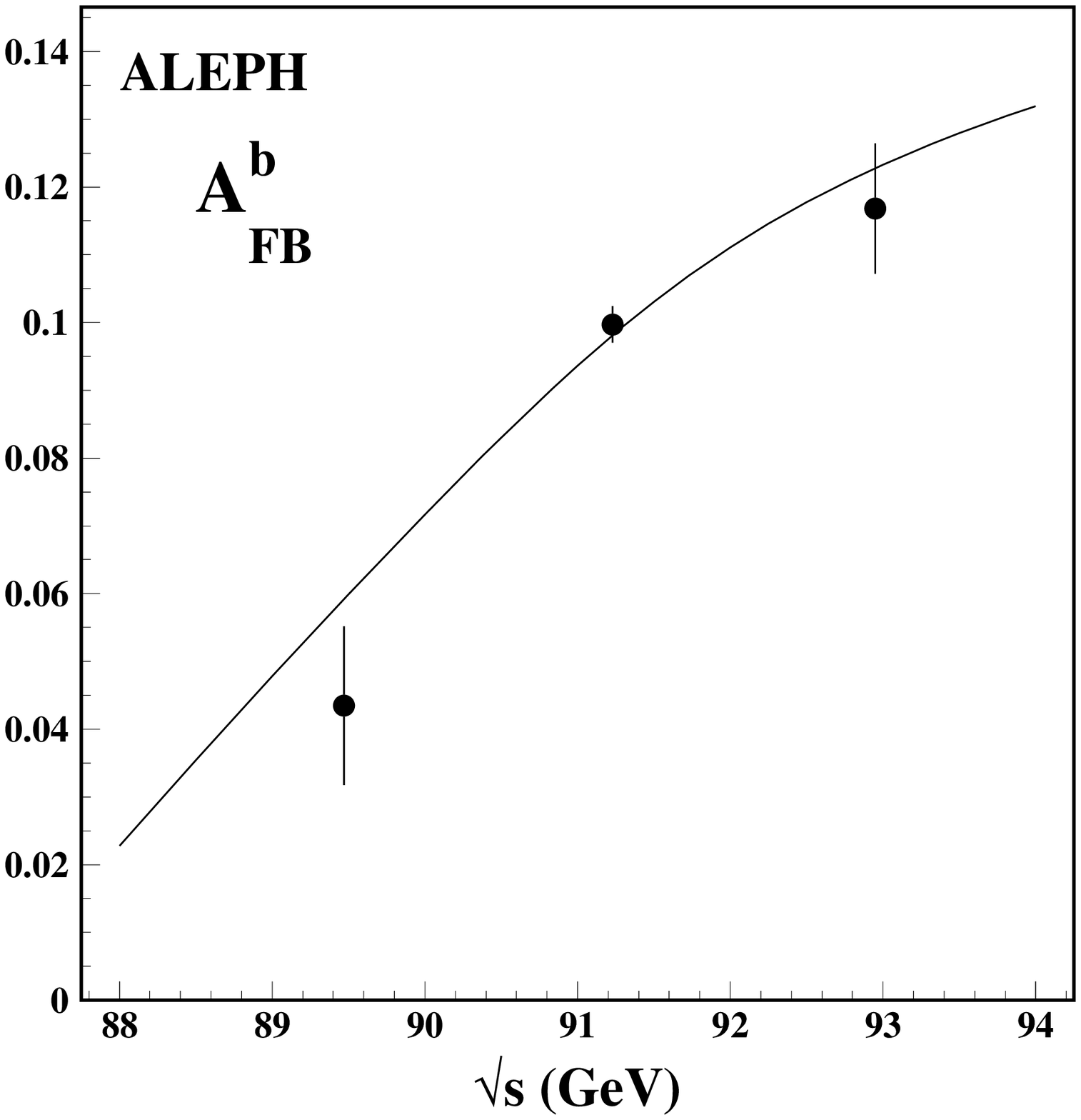,width=0.7 \textwidth}}\end{center}
\vspace{0.2 cm}
\caption{ The \protect$\protect\sqrt{s}\protect$ dependence of 
\protect$A^b_{\mathrm{FB}}\protect$. The errors shown are the statistical
errors and
the curve is the ZFITTER prediction, adjusted to reproduce the average fitted
asymmetry at $\sqrt{s} = M_Z$.}
\label{fig_result}
\end{figure}

The fit to Equation~(\ref{eq_afb}) is performed in three intervals
of collision energy, $\sqrt{s}$, with the results shown in Figure
\ref{fig_costh}. The asymmetries are then multiplied by the
QCD correction factor of $1.0034$ giving, for
the sample of events recorded closest to the $Z$ peak, the value:
\begin{eqnarray}
\nonumber                      
A^b_{\mathrm{FB}}(\sqrt{s}=91.232~{\rm GeV}) = 0.1000 \pm 0.0027(stat) \pm
0.0011(syst) \, .
\end{eqnarray} 
The corresponding values averaged over two off--peak energy ranges are
\begin{eqnarray}
\nonumber
A^b_{\mathrm{FB}}(\sqrt{s}=89.472~{\rm GeV}) = 0.0436 \pm 0.0119(stat) \, , \\ 
\nonumber
\vspace{0.3cm}
A^b_{\mathrm{FB}}(\sqrt{s}=92.950~{\rm GeV}) = 0.1172 \pm 0.0097(stat) \, . 
\end{eqnarray}
The variation of  $A^b_{\mathrm{FB}}$ with \mbox{centre-of-mass} energy is
shown in Figure~\ref{fig_result} and compared with the Standard Model
prediction. The three values are extrapolated to
$M_Z = 91.1874$ GeV using ZFITTER~\cite{ZFITTER},
giving the
combined value:
\begin{eqnarray}
\nonumber                      
A^b_{\mathrm{FB}}(\sqrt{s}=M_Z )  & = &
 0.0977 \, \pm \, 0.0025(stat)
\, \pm \, 0.0011(syst) \\
  & & + 0.103 \, (A^c_{\mathrm{FB}}-0.06513)-0.440 (R_b - 0.21585)\, .
\nonumber
\end{eqnarray}

In order to interpret the result in terms of the Standard Model,
a QED and $Z-\gamma$ interference correction
of 0.0038 is 
applied according to~\cite{qed}  to give the {\em pole asymmetry},
$A^{0,b}_{\rm FB}$, which is then iterated using BHM \cite{bhm}
until consistent values
for $A^{0,b}_{\rm FB}$, $A^{c}_{\rm FB}$ and $R_b$ are reached.
The result is
\begin{eqnarray*}
\nonumber                      
 A^{0,b}_{\rm FB} & = & 0.1009 \, \pm \, 0.0027\, \pm \, 0.0012
\end{eqnarray*}
using $A^{c}_{\rm FB} = 0.0623$ and $R_b = 0.21653$. The errors have
been obtained from iterating the value one standard deviation above
the measured $A^{b}_{\rm FB}$.
The corresponding weak mixing angle is
\begin{eqnarray}
\nonumber
{\rm sin^2} \theta_{\rm W}^{\rm eff} = 0.23193 \pm 0.00056\, .
\end{eqnarray}

This is to date the best $A^b_{\rm FB}$ measurement [2,26-34]
and provides the single most precise measurement of
${\rm sin^2} \theta_{\rm W}^{\rm eff}$ at LEP.
The value is in agreement with the present world
average value \cite{qed}.

\section*{Acknowledgements}
We wish to thank our colleagues from
the accelerator divisions for the successful operation of LEP.
It is also a pleasure to thank the technical personnel of the
collaborating institutions for their support in constructing
and maintaining the ALEPH experiment. Those of the
collaboration from non-member states thank CERN for its
hospitality.

\clearpage

\input thebib.tex

\end{document}

%% file: authors.tex
\pagestyle{empty}
\newpage
\small
%
\newlength{\saveparskip}
\newlength{\savetextheight}
\newlength{\savetopmargin}
\newlength{\savetextwidth}
\newlength{\saveoddsidemargin}
\newlength{\savetopsep}
\setlength{\saveparskip}{\parskip}
\setlength{\savetextheight}{\textheight}
\setlength{\savetopmargin}{\topmargin}
\setlength{\savetextwidth}{\textwidth}
\setlength{\saveoddsidemargin}{\oddsidemargin}
\setlength{\savetopsep}{\topsep}
%
%
\setlength{\parskip}{0.0cm}
\setlength{\textheight}{25.0cm}
\setlength{\topmargin}{-1.5cm}
\setlength{\textwidth}{16 cm}
\setlength{\oddsidemargin}{-0.0cm}
\setlength{\topsep}{1mm}
\pretolerance=10000
\centerline{\large\bf The ALEPH Collaboration}
\footnotesize
\vspace{0.5cm}
{\raggedbottom
\begin{sloppypar}
\samepage\noindent
A.~Heister,
S.~Schael
\nopagebreak
\begin{center}
\parbox{15.5cm}{\sl\samepage
Physikalisches Institut das RWTH-Aachen, D-52056 Aachen, Germany}
\end{center}\end{sloppypar}
\vspace{2mm}
\begin{sloppypar}
\noindent
R.~Barate,
I.~De~Bonis,
D.~Decamp,
C.~Goy,
\mbox{J.-P.~Lees},
E.~Merle,
\mbox{M.-N.~Minard},
B.~Pietrzyk
\nopagebreak
\begin{center}
\parbox{15.5cm}{\sl\samepage
Laboratoire de Physique des Particules (LAPP), IN$^{2}$P$^{3}$-CNRS,
F-74019 Annecy-le-Vieux Cedex, France}
\end{center}\end{sloppypar}
\vspace{2mm}
\begin{sloppypar}
\noindent
S.~Bravo,
M.P.~Casado,
M.~Chmeissani,
J.M.~Crespo,
E.~Fernandez,
\mbox{M.~Fernandez-Bosman},
Ll.~Garrido,$^{15}$
E.~Graug\'{e}s,
M.~Martinez,
G.~Merino,
R.~Miquel,$^{27}$
Ll.M.~Mir,$^{27}$
A.~Pacheco,
H.~Ruiz
\nopagebreak
\begin{center}
\parbox{15.5cm}{\sl\samepage
Institut de F\'{i}sica d'Altes Energies, Universitat Aut\`{o}noma
de Barcelona, E-08193 Bellaterra (Barcelona), Spain$^{7}$}
\end{center}\end{sloppypar}
\vspace{2mm}
\begin{sloppypar}
\noindent
A.~Colaleo,
D.~Creanza,
M.~de~Palma,
G.~Iaselli,
G.~Maggi,
M.~Maggi,
S.~Nuzzo,
A.~Ranieri,
G.~Raso,$^{23}$
F.~Ruggieri,
G.~Selvaggi,
L.~Silvestris,
P.~Tempesta,
A.~Tricomi,$^{3}$
G.~Zito
\nopagebreak
\begin{center}
\parbox{15.5cm}{\sl\samepage
Dipartimento di Fisica, INFN Sezione di Bari, I-70126
Bari, Italy}
\end{center}\end{sloppypar}
\vspace{2mm}
\begin{sloppypar}
\noindent
X.~Huang,
J.~Lin,
Q. Ouyang,
T.~Wang,
Y.~Xie,
R.~Xu,
S.~Xue,
J.~Zhang,
L.~Zhang,
W.~Zhao
\nopagebreak
\begin{center}
\parbox{15.5cm}{\sl\samepage
Institute of High Energy Physics, Academia Sinica, Beijing, The People's
Republic of China$^{8}$}
\end{center}\end{sloppypar}
\vspace{2mm}
\begin{sloppypar}
\noindent
D.~Abbaneo,
P.~Azzurri,
G.~Boix,$^{6}$
O.~Buchm\"uller,
M.~Cattaneo,
F.~Cerutti,
B.~Clerbaux,
G.~Dissertori,
H.~Drevermann,
R.W.~Forty,
M.~Frank,
T.C.~Greening,$^{29}$
J.B.~Hansen,
J.~Harvey,
P.~Janot,
B.~Jost,
M.~Kado,
P.~Mato,
A.~Moutoussi,
F.~Ranjard,
L.~Rolandi,
D.~Schlatter,
O.~Schneider,$^{2}$
W.~Tejessy,
F.~Teubert,
E.~Tournefier,$^{25}$
J.~Ward
\nopagebreak
\begin{center}
\parbox{15.5cm}{\sl\samepage
European Laboratory for Particle Physics (CERN), CH-1211 Geneva 23,
Switzerland}
\end{center}\end{sloppypar}
\vspace{2mm}
\begin{sloppypar}
\noindent
Z.~Ajaltouni,
F.~Badaud,
A.~Falvard,$^{22}$
P.~Gay,
P.~Henrard,
J.~Jousset,
B.~Michel,
S.~Monteil,
\mbox{J-C.~Montret},
D.~Pallin,
P.~Perret,
F.~Podlyski
\nopagebreak
\begin{center}
\parbox{15.5cm}{\sl\samepage
Laboratoire de Physique Corpusculaire, Universit\'e Blaise Pascal,
IN$^{2}$P$^{3}$-CNRS, Clermont-Ferrand, F-63177 Aubi\`{e}re, France}
\end{center}\end{sloppypar}
\vspace{2mm}
\begin{sloppypar}
\noindent
J.D.~Hansen,
J.R.~Hansen,
P.H.~Hansen,
B.S.~Nilsson,
A.~W\"a\"an\"anen
\begin{center}
\parbox{15.5cm}{\sl\samepage
Niels Bohr Institute, DK-2100 Copenhagen, Denmark$^{9}$}
\end{center}\end{sloppypar}
\vspace{2mm}
\begin{sloppypar}
\noindent
A.~Kyriakis,
C.~Markou,
E.~Simopoulou,
A.~Vayaki,
K.~Zachariadou
\nopagebreak
\begin{center}
\parbox{15.5cm}{\sl\samepage
Nuclear Research Center Demokritos (NRCD), GR-15310 Attiki, Greece}
\end{center}\end{sloppypar}
\vspace{2mm}
\begin{sloppypar}
\noindent
A.~Blondel,$^{12}$
G.~Bonneaud,
\mbox{J.-C.~Brient},
A.~Roug\'{e},
M.~Rumpf,
M.~Swynghedauw,
M.~Verderi,
\linebreak
H.~Videau
\nopagebreak
\begin{center}
\parbox{15.5cm}{\sl\samepage
Laboratoire de Physique Nucl\'eaire et des Hautes Energies, Ecole
Polytechnique, IN$^{2}$P$^{3}$-CNRS, \mbox{F-91128} Palaiseau Cedex, France}
\end{center}\end{sloppypar}
\vspace{2mm}
\begin{sloppypar}
\noindent
V.~Ciulli,
E.~Focardi,
G.~Parrini
\nopagebreak
\begin{center}
\parbox{15.5cm}{\sl\samepage
Dipartimento di Fisica, Universit\`a di Firenze, INFN Sezione di Firenze,
I-50125 Firenze, Italy}
\end{center}\end{sloppypar}
\vspace{2mm}
\begin{sloppypar}
\noindent
A.~Antonelli,
M.~Antonelli,
G.~Bencivenni,
G.~Bologna,$^{4}$
F.~Bossi,
P.~Campana,
G.~Capon,
V.~Chiarella,
P.~Laurelli,
G.~Mannocchi,$^{5}$
F.~Murtas,
G.P.~Murtas,
L.~Passalacqua,
\mbox{M.~Pepe-Altarelli},$^{24}$
P.~Spagnolo
\nopagebreak
\begin{center}
\parbox{15.5cm}{\sl\samepage
Laboratori Nazionali dell'INFN (LNF-INFN), I-00044 Frascati, Italy}
\end{center}\end{sloppypar}
\vspace{2mm}
\begin{sloppypar}
\noindent
A.W. Halley,
J.G.~Lynch,
P.~Negus,
V.~O'Shea,
C.~Raine,
A.S.~Thompson
\nopagebreak
\begin{center}
\parbox{15.5cm}{\sl\samepage
Department of Physics and Astronomy, University of Glasgow, Glasgow G12
8QQ,United Kingdom$^{10}$}
\end{center}\end{sloppypar}
\vspace{2mm}
\begin{sloppypar}
\noindent
S.~Wasserbaech
\nopagebreak
\begin{center}
\parbox{15.5cm}{\sl\samepage
Department of Physics, Haverford College, Haverford, PA 19041-1392, U.S.A.}
\end{center}\end{sloppypar}
\vspace{2mm}
\begin{sloppypar}
\noindent
R.~Cavanaugh,
S.~Dhamotharan,
C.~Geweniger,
P.~Hanke,
G.~Hansper,
V.~Hepp,
E.E.~Kluge,
A.~Putzer,
J.~Sommer,
K.~Tittel,
S.~Werner,$^{19}$
M.~Wunsch$^{19}$
\nopagebreak
\begin{center}
\parbox{15.5cm}{\sl\samepage
Kirchhoff-Institut f\"ur Physik, Universit\"at Heidelberg, D-69120
Heidelberg, Germany$^{16}$}
\end{center}\end{sloppypar}
\vspace{2mm}
\begin{sloppypar}
\noindent
R.~Beuselinck,
D.M.~Binnie,
W.~Cameron,
P.J.~Dornan,
M.~Girone,$^{1}$
N.~Marinelli,
J.K.~Sedgbeer,
J.C.~Thompson$^{14}$
\nopagebreak
\begin{center}
\parbox{15.5cm}{\sl\samepage
Department of Physics, Imperial College, London SW7 2BZ,
United Kingdom$^{10}$}
\end{center}\end{sloppypar}
\vspace{2mm}
\begin{sloppypar}
\noindent
V.M.~Ghete,
P.~Girtler,
E.~Kneringer,
D.~Kuhn,
G.~Rudolph
\nopagebreak
\begin{center}
\parbox{15.5cm}{\sl\samepage
Institut f\"ur Experimentalphysik, Universit\"at Innsbruck, A-6020
Innsbruck, Austria$^{18}$}
\end{center}\end{sloppypar}
\vspace{2mm}
\begin{sloppypar}
\noindent
E.~Bouhova-Thacker,
C.K.~Bowdery,
A.J.~Finch,
F.~Foster,
G.~Hughes,
R.W.L.~Jones,$^{1}$
M.R.~Pearson,
N.A.~Robertson
\nopagebreak
\begin{center}
\parbox{15.5cm}{\sl\samepage
Department of Physics, University of Lancaster, Lancaster LA1 4YB,
United Kingdom$^{10}$}
\end{center}\end{sloppypar}
\vspace{2mm}
\begin{sloppypar}
\noindent
I.~Giehl,
K.~Jakobs,
K.~Kleinknecht,
G.~Quast,
B.~Renk,
E.~Rohne,
\mbox{H.-G.~Sander},
H.~Wachsmuth,
C.~Zeitnitz
\nopagebreak
\begin{center}
\parbox{15.5cm}{\sl\samepage
Institut f\"ur Physik, Universit\"at Mainz, D-55099 Mainz, Germany$^{16}$}
\end{center}\end{sloppypar}
\vspace{2mm}
\begin{sloppypar}
\noindent
A.~Bonissent,
J.~Carr,
P.~Coyle,
O.~Leroy,
P.~Payre,
D.~Rousseau,
M.~Talby
\nopagebreak
\begin{center}
\parbox{15.5cm}{\sl\samepage
Centre de Physique des Particules, Universit\'e de la M\'editerran\'ee,
IN$^{2}$P$^{3}$-CNRS, F-13288 Marseille, France}
\end{center}\end{sloppypar}
\vspace{2mm}
\begin{sloppypar}
\noindent
M.~Aleppo,
F.~Ragusa
\nopagebreak
\begin{center}
\parbox{15.5cm}{\sl\samepage
Dipartimento di Fisica, Universit\`a di Milano e INFN Sezione di Milano,
I-20133 Milano, Italy}
\end{center}\end{sloppypar}
\vspace{2mm}
\begin{sloppypar}
\noindent
A.~David,
H.~Dietl,
G.~Ganis,$^{26}$
K.~H\"uttmann,
G.~L\"utjens,
C.~Mannert,
W.~M\"anner,
\mbox{H.-G.~Moser},
R.~Settles,
H.~Stenzel,
W.~Wiedenmann,
G.~Wolf
\nopagebreak
\begin{center}
\parbox{15.5cm}{\sl\samepage
Max-Planck-Institut f\"ur Physik, Werner-Heisenberg-Institut,
D-80805 M\"unchen, Germany\footnotemark[16]}
\end{center}\end{sloppypar}
\vspace{2mm}
\begin{sloppypar}
\noindent
J.~Boucrot,$^{1}$
O.~Callot,
M.~Davier,
L.~Duflot,
\mbox{J.-F.~Grivaz},
Ph.~Heusse,
A.~Jacholkowska,$^{22}$
J.~Lefran\c{c}ois,
\mbox{J.-J.~Veillet},
I.~Videau,
C.~Yuan
\nopagebreak
\begin{center}
\parbox{15.5cm}{\sl\samepage
Laboratoire de l'Acc\'el\'erateur Lin\'eaire, Universit\'e de Paris-Sud,
IN$^{2}$P$^{3}$-CNRS, F-91898 Orsay Cedex, France}
\end{center}\end{sloppypar}
\vspace{2mm}
\begin{sloppypar}
\noindent
G.~Bagliesi,
T.~Boccali,
G.~Calderini,
L.~Fo\`{a},
A.~Giammanco,
A.~Giassi,
F.~Ligabue,
A.~Messineo,
F.~Palla,
G.~Sanguinetti,
A.~Sciab\`a,
G.~Sguazzoni,
R.~Tenchini,$^{1}$
A.~Venturi,
P.G.~Verdini
\samepage
\begin{center}
\parbox{15.5cm}{\sl\samepage
Dipartimento di Fisica dell'Universit\`a, INFN Sezione di Pisa,
e Scuola Normale Superiore, I-56010 Pisa, Italy}
\end{center}\end{sloppypar}
\vspace{2mm}
\begin{sloppypar}
\noindent
G.A.~Blair,
G.~Cowan,
M.G.~Green,
T.~Medcalf,
A.~Misiejuk,
J.A.~Strong,
\mbox{P.~Teixeira-Dias},
\mbox{J.H.~von~Wimmersperg-Toeller}
\nopagebreak
\begin{center}
\parbox{15.5cm}{\sl\samepage
Department of Physics, Royal Holloway \& Bedford New College,
University of London, Egham, Surrey TW20 OEX, United Kingdom$^{10}$}
\end{center}\end{sloppypar}
\vspace{2mm}
\begin{sloppypar}
\noindent
R.W.~Clifft,
T.R.~Edgecock,
P.R.~Norton,
I.R.~Tomalin
\nopagebreak
\begin{center}
\parbox{15.5cm}{\sl\samepage
Particle Physics Dept., Rutherford Appleton Laboratory,
Chilton, Didcot, Oxon OX11 OQX, United Kingdom$^{10}$}
\end{center}\end{sloppypar}
\vspace{2mm}
\begin{sloppypar}
\noindent
\mbox{B.~Bloch-Devaux},$^{1}$
P.~Colas,
S.~Emery,
W.~Kozanecki,
E.~Lan\c{c}on,
\mbox{M.-C.~Lemaire},
E.~Locci,
P.~Perez,
J.~Rander,
\mbox{J.-F.~Renardy},
A.~Roussarie,
\mbox{J.-P.~Schuller},
J.~Schwindling,
A.~Trabelsi,$^{21}$
B.~Vallage
\nopagebreak
\begin{center}
\parbox{15.5cm}{\sl\samepage
CEA, DAPNIA/Service de Physique des Particules,
CE-Saclay, F-91191 Gif-sur-Yvette Cedex, France$^{17}$}
\end{center}\end{sloppypar}
\vspace{2mm}
\begin{sloppypar}
\noindent
N.~Konstantinidis,
A.M.~Litke,
G.~Taylor
\nopagebreak
\begin{center}
\parbox{15.5cm}{\sl\samepage
Institute for Particle Physics, University of California at
Santa Cruz, Santa Cruz, CA 95064, USA$^{13}$}
\end{center}\end{sloppypar}
\vspace{2mm}
\begin{sloppypar}
\noindent
C.N.~Booth,
S.~Cartwright,
F.~Combley,$^{4}$
M.~Lehto,
L.F.~Thompson
\nopagebreak
\begin{center}
\parbox{15.5cm}{\sl\samepage
Department of Physics, University of Sheffield, Sheffield S3 7RH,
United Kingdom$^{10}$}
\end{center}\end{sloppypar}
\vspace{2mm}
\begin{sloppypar}
\noindent
K.~Affholderbach,$^{28}$
A.~B\"ohrer,
S.~Brandt,
C.~Grupen,
A.~Ngac,
G.~Prange,
U.~Sieler
\nopagebreak
\begin{center}
\parbox{15.5cm}{\sl\samepage
Fachbereich Physik, Universit\"at Siegen, D-57068 Siegen,
 Germany$^{16}$}
\end{center}\end{sloppypar}
\vspace{2mm}
\begin{sloppypar}
\noindent
G.~Giannini
\nopagebreak
\begin{center}
\parbox{15.5cm}{\sl\samepage
Dipartimento di Fisica, Universit\`a di Trieste e INFN Sezione di Trieste,
I-34127 Trieste, Italy}
\end{center}\end{sloppypar}
\vspace{2mm}
\begin{sloppypar}
\noindent
J.~Rothberg
\nopagebreak
\begin{center}
\parbox{15.5cm}{\sl\samepage
Experimental Elementary Particle Physics, University of Washington, Seattle, 
WA 98195 U.S.A.}
\end{center}\end{sloppypar}
\vspace{2mm}
\begin{sloppypar}
\noindent
S.R.~Armstrong,
K.~Cranmer,
P.~Elmer,
D.P.S.~Ferguson,
Y.~Gao,$^{20}$
S.~Gonz\'{a}lez,
O.J.~Hayes,
H.~Hu,
S.~Jin,
J.~Kile,
P.A.~McNamara III,
J.~Nielsen,
W.~Orejudos,
Y.B.~Pan,
Y.~Saadi,
I.J.~Scott,
J.~Walsh,
Sau~Lan~Wu,
X.~Wu,
G.~Zobernig
\nopagebreak
\begin{center}
\parbox{15.5cm}{\sl\samepage
Department of Physics, University of Wisconsin, Madison, WI 53706,
USA$^{11}$}
\end{center}\end{sloppypar}
}
\footnotetext[1]{Also at CERN, 1211 Geneva 23, Switzerland.}
\footnotetext[2]{Now at Universit\'e de Lausanne, 1015 Lausanne, Switzerland.}
\footnotetext[3]{Also at Dipartimento di Fisica di Catania and INFN Sezione di
 Catania, 95129 Catania, Italy.}
\footnotetext[4]{Deceased.}
\footnotetext[5]{Also Istituto di Cosmo-Geofisica del C.N.R., Torino,
Italy.}
\footnotetext[6]{Supported by the Commission of the European Communities,
contract ERBFMBICT982894.}
\footnotetext[7]{Supported by CICYT, Spain.}
\footnotetext[8]{Supported by the National Science Foundation of China.}
\footnotetext[9]{Supported by the Danish Natural Science Research Council.}
\footnotetext[10]{Supported by the UK Particle Physics and Astronomy Research
Council.}
\footnotetext[11]{Supported by the US Department of Energy, grant
DE-FG0295-ER40896.}
\footnotetext[12]{Now at Departement de Physique Corpusculaire, Universit\'e de
Gen\`eve, 1211 Gen\`eve 4, Switzerland.}
\footnotetext[13]{Supported by the US Department of Energy,
grant DE-FG03-92ER40689.}
\footnotetext[14]{Also at Rutherford Appleton Laboratory, Chilton, Didcot, UK.}
\footnotetext[15]{Permanent address: Universitat de Barcelona, 08208 Barcelona,
Spain.}
\footnotetext[16]{Supported by the Bundesministerium f\"ur Bildung,
Wissenschaft, Forschung und Technologie, Germany.}
\footnotetext[17]{Supported by the Direction des Sciences de la
Mati\`ere, C.E.A.}
\footnotetext[18]{Supported by the Austrian Ministry for Science and Transport.}
\footnotetext[19]{Now at SAP AG, 69185 Walldorf, Germany.}
\footnotetext[20]{Also at Department of Physics, Tsinghua University, Beijing, The People's Republic of China.}
\footnotetext[21]{Now at D\'epartement de Physique, Facult\'e des Sciences de Tunis, 1060 Le Belv\'ed\`ere, Tunisia.}
\footnotetext[22]{Now at Groupe d' Astroparticules de Montpellier, Universit\'e de Montpellier II, 34095 Montpellier, France.}
\footnotetext[23]{Also at Dipartimento di Fisica e Tecnologie Relative, Universit\`a di Palermo, Palermo, Italy.}
\footnotetext[24]{Now at CERN, 1211 Geneva 23, Switzerland.}
\footnotetext[25]{Now at ISN, Institut des Sciences Nucl\'eaires, 53 Av. des Martyrs, 38026 Grenoble, France.} 
\footnotetext[26]{Now at INFN Sezione di Roma II, Dipartimento di Fisica, Universit\'a di Roma Tor Vergata, 00133 Roma, Italy.} 
\footnotetext[27]{Now at LBNL, Berkeley, CA 94720, U.S.A.}
\footnotetext[28]{Now at Skyguide, Swissair Navigation Services, Geneva, Switzerland.}
\footnotetext[29]{now at Honeywell, Phoenix AZ, U.S.A.}
%
\setlength{\parskip}{\saveparskip}
\setlength{\textheight}{\savetextheight}
\setlength{\topmargin}{\savetopmargin}
\setlength{\textwidth}{\savetextwidth}
\setlength{\oddsidemargin}{\saveoddsidemargin}
\setlength{\topsep}{\savetopsep}
\normalsize
\newpage
\pagestyle{plain}
\setcounter{page}{1}
